\documentclass[pdflatex,sn-nature]{sn-jnl}

\usepackage{graphicx}%
\usepackage{multirow}%
\usepackage{amsmath,amssymb,amsfonts}%
\usepackage{amsthm}%
\usepackage{mathrsfs}%
\usepackage[title]{appendix}%
\usepackage{xcolor}%
\usepackage{textcomp}%
\usepackage{manyfoot}%
\usepackage{booktabs}%
\usepackage{algorithm}%
\usepackage{algorithmicx}%
\usepackage{algpseudocode}%
\usepackage{listings}%

\usepackage{bm}
\usepackage[capitalize]{cleveref}
\usepackage{braket}
\usepackage{siunitx}
\usepackage{mathtools}
\usepackage{mhchem}
\usepackage{ulem}
\newcommand{\MyTr}[1]{{\mathrm{Tr}\,\left[ {#1} \right]}}
\newcommand{\tov}[1][]{{\tilde{t}_{\mathrm{sw}}^{#1}}}
\newcommand{\tobs}[1][]{{t_{\mathrm{obs}}^{#1}}}
\newcommand{\Nobs}{{N_{\mathrm{obs}}}}

\newcommand{\PSD}[2][]{{\mathcal{P}^{#1}_{#2}}}
\newcommand{\SPSD}[2][]{{\mathcal{S}^{#1}_{#2}}}
\newcommand{\BPSD}[2][]{{\mathcal{B}^{#1}_{#2}}}

\theoremstyle{thmstyleone}%
\theoremstyle{thmstyletwo}%

\theoremstyle{thmstylethree}%

\begin{document}

\title[Article Title]{Hybrid-spin decoupling for noise-resilient DC quantum sensing}

\author*[1]{\fnm{So} \sur{Chigusa}}\email{schigusa@mit.edu}
\equalcont{These authors contributed equally to this work.}

\author[2,3]{\fnm{Masashi} \sur{Hazumi}}

\author*[4]{\fnm{Ernst David} \sur{Herbschleb}}\email{herbschleb@dia.kuicr.kyoto-u.ac.jp}
\equalcont{These authors contributed equally to this work.}

\author[5]{\fnm{Yuichiro} \sur{Matsuzaki}}

\author[4,6,7]{\fnm{Norikazu} \sur{Mizuochi}}

\author[6,8]{\fnm{Kazunori} \sur{Nakayama}}

\affil[1]{\orgdiv{Center for Theoretical Physics - a Leinweber Institute}, \orgname{Massachusetts Institute of Technology}, \orgaddress{\street{77 Massachusetts Avenue}, \city{Cambridge}, \postcode{02139}, \state{MA}, \country{USA}}}

\affil[2]{\orgdiv{Institute of Particle and Nuclear Studies (IPNS)}, \orgname{KEK}, \orgaddress{\city{Tsukuba}, \postcode{305-0801}, \state{Ibaraki}, \country{Japan}}}

\affil[3]{\orgdiv{Department of Physics and Center for High Energy and High Field Physics (CHiP)}, \orgname{National Central University}, \orgaddress{\city{Taoyuan City}, \state{Zhongli District}, \country{Taiwan}}}

\affil[4]{\orgdiv{Institute for Chemical Research}, \orgname{Kyoto University}, \orgaddress{\street{Gokasho}, \city{Uji}, \postcode{611-0011}, \state{Kyoto}, \country{Japan}}}

\affil[5]{\orgdiv{Department of Electrical, Electronic, and Communication Engineering}, \orgname{Chuo University}, \orgaddress{\street{Kasuga}, \city{Bunkyo}, \postcode{112-8551}, \state{Tokyo}, \country{Japan}}}

\affil[6]{\orgdiv{International Center for Quantum-field Measurement Systems for Studies of the Universe and Particles (QUP)}, \orgname{KEK}, \orgaddress{\street{1-1 Oho}, \city{Tsukuba}, \postcode{305-0801}, \state{Ibaraki}, \country{Japan}}}

\affil[7]{\orgdiv{Center for Spintronics Research Network}, \orgname{Kyoto University}, \orgaddress{\city{Uji}, \postcode{611-0011}, \state{Kyoto}, \country{Japan}}}

\affil[8]{\orgdiv{Department of Physics}, \orgname{Tohoku University}, \orgaddress{\city{Sendai}, \postcode{980-8578}, \state{Miyagi}, \country{Japan}}}

\abstract{The excellent sensitivities of quantum sensors are a double-edged sword: minuscule quantities can be observed, but any undesired signal acts as noise. This is challenging when detecting quantities that are obscured by such noise. Decoupling sequences improve coherence times and hence sensitivities, though only AC signals in narrow frequency bands are distinguishable. Alternatively, comagnetometers operate gaseous spin mixtures at high temperatures in the self-compensating regime to counteract slowly varying noise. These are applied with great success in various exotic spin-interaction searches. Here, we propose a method that decouples specific DC fields from DC and AC magnetic noise. It requires any spin cluster where the effect on each individual spin is different for the target field and local magnetic fields, which allows for a different approach compared to comagnetometers. The presented method has several key advantages, including an orders-of-magnitude increase in noise frequencies to which we are resistant. We explore electron-spin nuclear-spin pairs in nitrogen-vacancy centres in diamond, with a focus on their merit for light dark-matter searches. Other applications include gradient sensing, quantum memory, and gyroscopes.}



\maketitle



Quantum sensors have superb sensitivities that enable applications in many fields. An example is the nitrogen-vacancy (NV) centre~\cite{Degen:2017}, which has outstanding quantum properties even under ambient conditions, such as long electron spin coherence times~\cite{Herbschleb:2019}. Moreover, the nitrogen atom and carbon-$13$ atoms within the diamond lattice allow for hybrid protocols that include both electron spins and nuclear spins.

There are two main ways in which the nuclear spin is combined with the electron spin in sensing protocols~\cite{Kenny:2025}. The first involves temporary storage of the electron spin state in the nuclear spin state to overcome the limited dephasing time of the electron spin. Examples include a delayed echo to measure specific nuclear spins around the NV centre~\cite{Wang:2017}, an increased gradient evolution time to sharpen filter linewidths~\cite{Ajoy:2015}, and an extension of the dephasing time~\cite{Chen:2023}.
The second way is the auxiliary use of the nuclear spin for enhanced spin readout techniques~\cite{Dreau:2013}. 
A similar principle holds when the electron spin is essentially ancillary for reading out and initialising the nuclear spin itself~\cite{Xie:2021, Burgler:2023}.

Both ways rely on the nuclear spin to overcome shortcomings of the electron spin: a limited dephasing time and imperfect readout. However, there are methods that principally require both electron and nuclear spins. The first demonstration is noble gas–alkali metal comagnetometry~\cite{Kornack:2002}, which generally operates with an overlapping gaseous K-$^3$He spin ensemble mixture in a heated spherical cell, for example with a $20$ mm diameter. An external magnetic field brings it into its self-compensating regime, allowing it to adiabatically follow slowly-changing magnetic fields (timescale of hundreds of milliseconds~\cite{Kornack:2002}), making it immune to such low-frequency noise. It is a challenging field, as e.g. a high density of the noble gas is preferred for operation at the equilibrium compensation point~\cite{Klinger:2023}, while a low density is preferred because collisions between the atoms result in uncompensated noise~\cite{Padniuk:2024}. Such comagnetometers are employed for axion dark-matter searches, recently improving laboratory constraints on various axion couplings~\cite{Gavilan:2025}.

With just a single kind of spin, dynamical decoupling sequences, such as Carr-Purcell-Meiboom-Gill~\cite{MeiboomGill:1958}, also allow one to reduce the effect of low-frequency noise. By applying a sequence of $\pi$-pulses, the signal is effectively filtered, and the spin becomes only sensitive to a specific frequency. By increasing the number of pulses, the filter function becomes sharper, and this frequency increases. As such, the coherence time related to the sequence becomes longer with an upper limit related to $T_1$~\cite{BarGill:2013}. Lowering the temperature also increases $T_1$, and hence longer coherence times become possible with more pulses.

In this paper, we propose a hybrid spin protocol, here investigated for solid-state diamond, to cancel magnetic noise, which is, e.g., useful for quantum memories~\cite{Maurer:2012}. Using two different spins (for example an electron spin and a nuclear spin), any interaction that does not have the same relative effect on both spins as a local magnetic field can be detected accurately, since the magnetic noise can be cancelled. Examples are field gradients~\cite{Ilias:2024}, gyroscopes~\cite{Kornack:2005} useful for navigation~\cite{Fang:2012}, and potential exotic spin-interactions such as electric dipole moment~\cite{PhysRevLett.124.081803, PhysRevLett.123.143003, PhysRevA.100.022505}, spin-dependent fifth forces~\cite{Vasilakis:2008yn,Lee:2018, Almasi:2018cob}, spin-dependent gravitational interactions~\cite{Zhang:2023}, and breaking of CPT and/or Lorentz symmetries~\cite{Kornack:thesis, Brown:2010dt, Smiciklas:2011}.

There are five key advantages over comagnetometry. First, the DC quantity to detect is protected from low-frequency noise, which already significantly enlarges the frequency spectrum of reduced noise. Second, as long as fast high-fidelity gate operations are available, additional gates decouple the signal from common noise sources up to high frequencies, further increasing the noise spectrum and, moreover, elongating effective coherence times to up to $T_1$. Third, since low-temperature operation is possible which elongates $T_1$~\cite{BarGill:2013}, additional enhanced sensitivities are available. Fourth, applying the method with a single hybrid spin cluster unlocks the possibility for high spatial resolution sensing~\cite{degen:2008}. Fifth, in solid state, expanding the ensemble size for increasing the sensitivity requires significantly less space. In conclusion, a noise-resistant quantum sensor with exquisite broadband sensitivities for exotic fields is possible. Here, we investigate the potential of our technique for axion dark matter detection with NV centres.

In the following, first, we introduce the protocol. Then, we study its relaxation time, the effect of noise, and the advantages for dark-matter search. Finally, we look at future directions.


\section*{Protocol}
\label{sec:protocol}

\begin{figure}[h]
  \centering
  \hspace*{-2cm} 
  \includegraphics{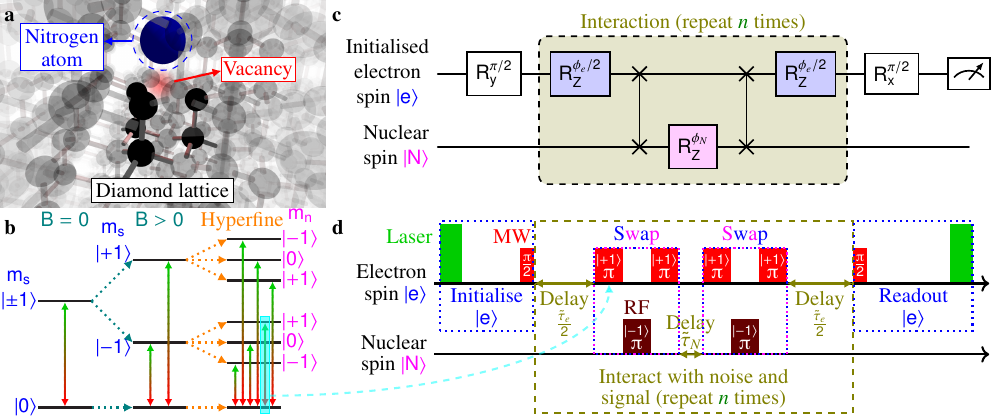}
  \caption{Protocol. \textbf{a} In a diamond lattice, a substitutional nitrogen atom and a neighbouring missing carbon atom form a nitrogen-vacancy (NV) centre. The negatively charged NV centre has an electron spin, while the nitrogen atom has a nuclear spin. \textbf{b} Energy level diagram for the electron spin of the NV centre only for simplicity (not to scale). Spin-1 systems have a splitting at zero field, while a magnetic field $B$ splits its $\ket{\pm1}$ states, and hyperfine coupling splits these further depending on the state of the nuclear spin. \textbf{c} Graph of the proposed protocol. Swap gates allow both spins to contribute to the sensing result. \textbf{d} Example pulse sequence to implement the graph from \textbf{c}. A green laser initialises the electron spin, and a microwave (MW) creates a superposition. After an initial interaction delay, the state is swapped to the nuclear spin, and later it is swapped back. Swaps are possible by using three controlled-not gates using the hyperfine splitting, an example is indicated with the cyan arrow for a transition in \textbf{b}. Finally, the final state of the electron spin is read with a MW and laser pulse. Not to scale, notably generally $\tilde{\tau}_e \ll \tilde{\tau}_N$.}
  \label{fig:protocol}
\end{figure}

Our protocol combines two different spins, for example, an electron spin of a nitrogen vacancy centre and the nuclear spin of its nitrogen (see Fig.~\ref{fig:protocol}a). Owing to the proximity of the spins, spin-dependent transitions are possible via the hyperfine interaction~\cite{Neumann:2010}, as illustrated by the energy level diagram for the electron spin in Fig.~\ref{fig:protocol}b. In the protocol as depicted in Fig.~\ref{fig:protocol}c, after initialising the electron spin into a superposition, the state is alternated between the two spin species by swap gates. Finally, the electron spin, which holds the final spin state, is measured. By properly choosing the ratio between the times the state spends in each spin, specific shared signals for the spins are cancelled from the state, while differences stand out. We call the optimal ratio the fine-tuning condition, which follows from
\begin{align}
  \gamma_e \tau_e + \gamma_N \tau_N=0,
  \label{eq:fine-tuning}
\end{align}
with $\gamma_e$ and $\gamma_N$ the gyromagnetic ratios of the electron and nuclear spins, respectively, and $\tau_e$ and $\tau_N$ the time the state spends within each spin. The following section shows the origin of this condition.

Figure~\ref{fig:protocol}d shows an example implementation of this protocol for NV centres. A laser pulse initialises the electron spin into the $\ket{0}$ state~\cite{Degen:2017}. Next, a microwave (MW) $\pi/2$-pulse creates a superposition. During a delay $\tilde{\tau}_e/2$, the electron spin interacts with the environment. Then, a swap gate, e.g. implemented by applying MW (for the electron spin) and radio frequency (RF; for the nuclear spin) pulses to specific energy level differences, moves the electron spin state to the nuclear spin state~\cite{Neumann:2010}. After a delay $\tilde{\tau}_N$, the state is switched back to the electron spin, which interacts for another $\tilde{\tau}_e/2$.
Note that the timescales in \cref{eq:fine-tuning} are related to these timescales as $\tau_e = n\tilde{\tau}_e$ and $\tau_N = n\tilde{\tau}_N$.
Finally, the resulting phase is projected onto the $\ket{0}$-$\ket{-1}$ basis, and it is read via a second laser pulse.

Essentially, a unit cell of the protocol would simply be: electron-spin-accumulation, swap, nuclear-spin accumulation. However, since repetition of this unit cell is investigated as well, while readout of the nuclear spin would happen via the electron spin, we choose the repeatable unit cell consisting of two swap gates with the electron-spin accumulation time divided over the start and the end. Repeating $n$ times results in noise cancellation over a larger range of frequencies, as explored in the following sections. As this is thus a decoupling sequence for which the use of two different spins is essential, we name it the hybrid-spin decoupling protocol. Interestingly, even though original decoupling sequences are only sensitive to AC fields with a specific frequency~\cite{MeiboomGill:1958}, the fields targeted by our protocol are DC.

In principle, any pair of spins with the desired properties for an application can be chosen. The advantage of the NV centre is that each includes both an electron spin and a nuclear spin, therefore operation of ensembles of NV centres can yield major improvements in sensitivity.

\section*{Relaxation time}
\label{sec:relaxation}

The relaxation time scales of the spins can be estimated using the protocol-dependent method used in, e.g., \cite{PhysRev.125.912,PhysRevLett.97.087601,PhysRevB.87.115122,doi:10.7566/JPSJ.89.054708}.
As a random noise model, we assume the interaction Hamiltonian
\begin{align}
  H_I(t) &= \lambda f(t) g(t) \sigma_z, \label{eq:H_I}
\end{align}
with $\lambda$ the strength of the noise, $f(t)$ the normalised random function characterised by two-point correlation functions
\begin{align}
  \Braket{f(t)f(0)} = e^{-|t|/\tau_c}, \label{eq:f2}
\end{align}
with $\Braket{\cdot}$ denoting the ensemble average.
For simplicity of calculation, we assume a Gaussian noise with zero average, $\Braket{f(t)} = 0$.
$g(t)$ is the filter function that represents the detection protocol.
For our hybrid-spin decoupling protocol, it is defined as
\newcommand{\gHD}{%
  g(t) = \sum_{k=0}^{n-1} \Bigg\{ \,
    & \gamma_e \Theta\left[\left(
        t-k\tilde{\tau}
      \right) \left(
        k\tilde{\tau}+\frac{\tilde{\tau}_e}{2}-t
      \right)\right] \notag \\
    & + \gamma_N \Theta\left[ \left(
        t-k\tilde{\tau}-\frac{\tilde{\tau}_e}{2}-\tov
      \right) \left(
        k\tilde{\tau}+\frac{\tilde{\tau}_e}{2}+\tilde{\tau}_N+\tov-t
      \right)\right] \notag \\
    & + \gamma_e \Theta\left[\left(
        t-k\tilde{\tau}-\frac{\tilde{\tau}_e}{2}-\tilde{\tau}_N-2\tov
      \right) \left(
        (k+1)\tilde{\tau}-t
      \right)\right]
  \Bigg\}
}
\begin{align}
  \gHD,
  \label{eq:g_HD}
\end{align}
where $\Theta$ is the Heaviside function, $\tov$ is the swap gate operation time, $\tilde{\tau} = \tilde{\tau}_N + \tilde{\tau}_e + 2\tov$ is the interaction unit time, and $n$ is the number of repetitions.
Neglecting overhead such as initialisation and readout, $\tau = n\tilde{\tau}$ approximates the time duration for the measurement.

\begin{figure}[h]
  \centering
  \hspace*{-2cm} 
  \includegraphics{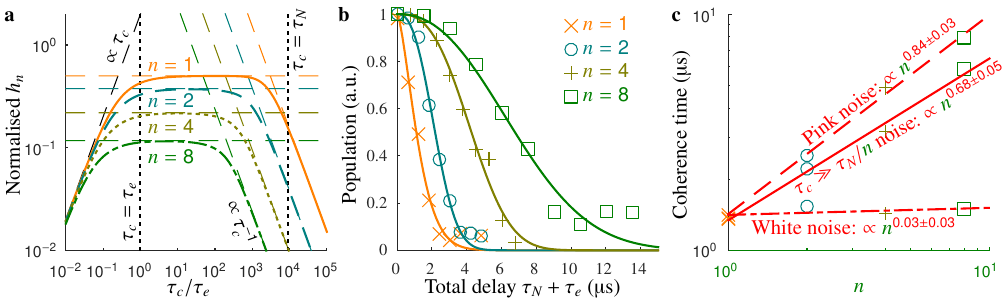}
  \caption{Coherence. \textbf{a} Normalised $h_n$ versus different correlation times $\tau_c$ of the noise, for various $n$. The gyromagnetic ratio of the electron spin and the $^{14}$N nuclear spin are used, and the fine-tuning condition of \cref{eq:fine-tuning} is applied. Dashed lines represent the asymptotic behaviours in the three regimes of $\tau_c$. \textbf{b} Simulated decays for $4$ different $n$s with pink noise. The magnitude of the noise is chosen to limit the coherence time within the simulation time window. \textbf{c} Dependence of the coherence time on $n$. Given \cref{eq:sigma_exp}, for noise following $\tau_c \gg \tilde{\tau}_N$, the coherence time is proportional to $n^{0.68}$ (red line). For white noise, which is most similar to $\tau_c \ll \tilde{\tau}_e$, the coherence time is nearly independent of $n$ (red dashed-dotted line). For pink noise, a mixture of regimes applies, which results in a more complex dependency (red dashed line).}
  \label{fig:coherence}
\end{figure}

Within this noise model, the evolution of the quantum state can be explicitly calculated in terms of an ensemble average (see \cref{sec:relaxation} for details).
Taking $\rho(t=0) = \ket{+}\bra{+}$ with $\ket{+}\equiv \frac{1}{\sqrt{2}}(\ket{g} + \ket{e})$ as an initial state with $\ket{g}$ and $\ket{e}$ being the ground and excited states of a qubit, the relevant relaxation is captured by the exponential decay of the transverse polarisation, $\Braket{\sigma_x}_\tau$, where $\Braket{\cdot}_t$ denotes the ensemble average of the expectation value of an operator at time $t$.
Parameterising the decay behaviour for the hybrid-spin decoupling protocol as $\Braket{\sigma_x}_\tau = e^{-h_n}$, the exponent $h_n$ depends on the hierarchy among the relevant timescales as
\begin{align}
  h_n \to \begin{dcases}
    2\lambda^2 \left(
      \gamma_N \tau_N + \gamma_e \tau_e
    \right)^2 + \mathcal{O}(\tau_c^{-1}), & (\tau_c \gg \tilde{\tau}_N) \\
    4\lambda^2 \gamma_N \tau_c \left(
      \gamma_N \tau_N + \gamma_e \tau_e
    \right) + \frac{(2n-1)\lambda^2 \gamma_e^2 \tau_e^2}{n^2}, & (\tilde{\tau}_N \gg \tau_c \gg \tilde{\tau}_e) \\
    4\lambda^2 \gamma_e^2 \tau_c \tau_e, & (\tilde{\tau}_e \gg \tau_c)
  \end{dcases}
  \label{eq:sigma_exp}
\end{align}
where $n \ll \tau_N / \tau_e$ is assumed.
See \cref{eq:h_n_HSCP} for the full expression of $h_n$.
In particular, under the condition \cref{eq:fine-tuning} with negligible $\tov$, noises with a long correlation time $\tau_c \gg \tilde{\tau}_N$ result in $\Braket{\sigma_x}_\tau \simeq \exp\left[-\frac{\lambda^2\gamma_N^2\tau_N^3}{3n^2\tau_c}\right]$, which mimics the behaviour under the Carr-Purcell (CP) protocol in \cref{eq:tau_c_infty}.
Similarly, the result for a short correlation time $\tau_c \ll \tilde{\tau}_e$ is the same as that of the CP protocol in \cref{eq:tau_c_0}.
Finally, the behaviour for an intermediate correlation time $\tilde{\tau}_N \gg \tau_c \gg \tilde{\tau}_e$ can be interpreted as the dephasing caused by DC-like noise sources on the electron spin (see \cref{eq:tau_c_infty} for comparison).
It is worth noting that, in this regime, the decoherence originating from the electron spin can be suppressed by increasing $n$, owing to a phenomenon known as qubit motion~\cite{averin2016suppression,matsuzaki2016hybrid}. However, the main advantage from increasing $n$ resides in the change of $\tilde{\tau}_e$ and $\tilde{\tau}_N$ compared to $\tau_c$ for the proposed method.
These results are visualised in Fig.~\ref{fig:coherence}a, where the $y$-axis is a normalised exponent, $h_n / (2\lambda^2\gamma_e^2\tau_e^2)$, which is the exponent for a Ramsey sequence [\cref{eq:tau_c_infty}].

From the above expression, it can be observed that the exponent reduces for a larger $n$, which would reduce the overall relaxation effects.
Although we considered a single noise source above, let us assume there are at least two noise sources. One of them has a short $\tau_c$ while the other has an intermediate $\tau_c$. In this case, by comparing these results with those for the Ramsey and Hahn-echo protocols discussed in \cref{sec:Ramsey}, we expect the decoherence behaviour
\def\decoherence{
  \Braket{\sigma_x}_\tau \propto \exp \left[
    -\frac{2n-1}{n^2} \left(
      \frac{\tau_e}{T_2^*}
    \right)^2 - \frac{\tau_e}{T_2}
  \right],
}
\begin{align}
  \decoherence
  \label{eq:decoherence}   
\end{align}
for a moderate choice of $n\sim \mathcal{O}(1)$, where $T_2^*$ and $T_2$ are comparable to $T_{2e}^*\sim \SI{1}{\mu s}$ and $T_{2e}\sim \SI{100}{\mu s}$ for the electron spin~\cite{Barry:2020,Mizuochi:2009,Wolf:2015}.
See \cref{ymadd} for details.
Similarly to the ordinary dynamic decoupling protocols, the coherence time becomes longer for a larger $n$ and asymptotes to $T_{2e}$ for $n\to \infty$.
Below, we use $T_2^* = \SI{1}{\mu s}$ for performance demonstration of the protocol, leaving the experimental determination of the coherence time for future work.

To study the effect on coherence time of our protocol, we simulate a simplified Hamiltonian of the NV centre (see \cref{sec:simulation} for details, and Fig.~\ref{fig:coherence}b for results). The noise added to the model is chosen to be large enough for the coherence time to be within the simulated time span, while $\gamma_N=-0.5\gamma_e$ for clearer visualisations later. For pink noise, the coherence time scales as $n^{0.8}$ (see Fig.~\ref{fig:coherence}c), thus showing increased coherence times for larger $n$. This is similar to what we found above; the difference we attribute to the mixture of the regimes of \cref{eq:sigma_exp} and to the simulated condition $n \sim \mathcal{O}(1)$.

\section*{Noise estimation}

Following the same strategy as in the previous section, stochastic magnetic noise effects can also be discussed.
In an experiment environment, such random noises can be sourced from, e.g., fluctuations of the external magnetic field and leakage from the magnetic shielding, which makes it hard to sense the signal field below the noise level.
We analyse the interaction Hamiltonian \cref{eq:H_I}, where the two-point correlation of the random function $f(t)$ is now related to the power spectral density of the noise,
\begin{align}
  P(\omega) \equiv \lambda^2 \int dt\, e^{i\omega t} \Braket{f(t) f(0)}.
  \label{eq:noise_PSD}
\end{align}
which is an even function of $\omega$ due to the time-translation symmetry.

\begin{figure}[!h]
  \centering
  \hspace*{-2cm} 
  \includegraphics{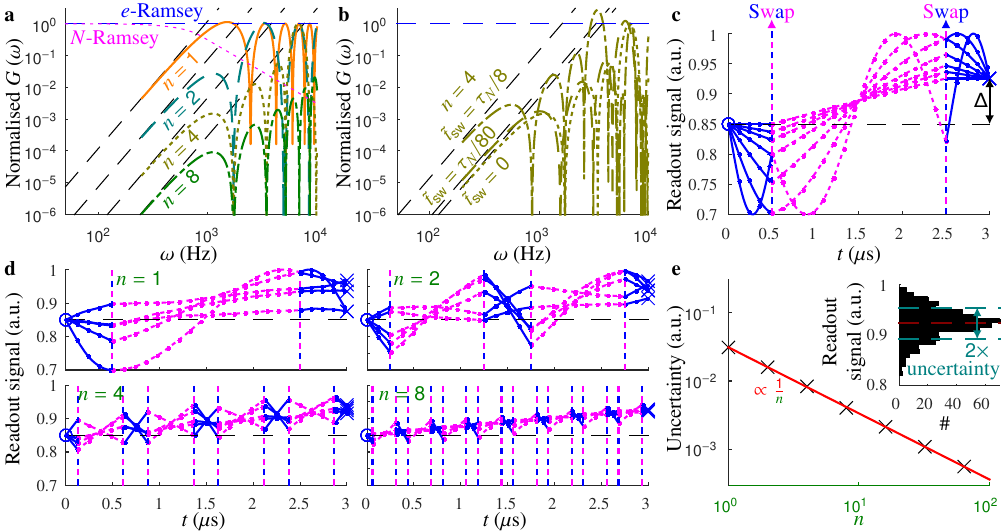}
  \caption{Noise. \textbf{a}. The noise contribution [calculated via \cref{eq:G}] in the hybrid-spin decoupling protocol as a function of the angular frequency $\omega$ for a white noise profile. The $y$-axis is normalized for demonstration. \textbf{b}. For $n=4$, the effect of several swap-gate lengths is calculated. For realistic lengths ($\tov\sim\tau_N/80$), the effect is negligible. \textbf{c} Simulation results for $n=1$ with a fixed difference in magnetic fields between the two spins, thus like a DC gradient field. The projection of the phase is plotted for each time during the sequence of Fig.~\ref{fig:protocol}d between the two $\pi/2$-pulses. For clarity, $\gamma_N=-0.5\gamma_e$ is used, so $\tau_N=2\tau_e$. The different lines are for different mutual DC magnetic fields at each spin, which could be DC noise. The DC gradient in magnetic field is detected. \textbf{d} Five example simulation results for four different repetitions $n$, with pink noise added. The noise is randomly set before a simulation, thus each simulation has different noise. \textbf{e} For each $n$, a simulation is run $500$ times, and the uncertainty in the readout signal (e.g. final data points in \textbf{d}) is plotted; the inset gives an example for finding this uncertainty for $n=1$. The fit shows a decrease in uncertainty with an increase in $n$.}
  \label{fig:noise}
\end{figure}

According to the detailed calculation shown in \cref{sec:noise}, the external noise leads to relaxation behaviour
\begin{align}
  \Braket{\sigma_x}_\tau = \exp\left[
    -4 \int_0^\infty \frac{d\omega}{2\pi}\, G(\omega)
  \right],
  \label{eq:relaxation_noise}
\end{align}
with
\begin{align}
  G(\omega) = P(\omega) \int_0^\tau dt_1\,dt_2\, e^{-i\omega(t_1-t_2)} g(t_1) g(t_2) .
  \label{eq:G}
\end{align}
The low-frequency expansion of the above equation for the hybrid-spin decoupling protocol shows the possible noise cancellation.
Indeed, the leading contribution is provided by
\newcommand{\GHDLF}{%
  G(\omega) \simeq P(\omega) \left[
    (\gamma_N \tau_N + \gamma_e \tau_e)^2
    + (\gamma_N \tau_N + \gamma_e \tau_e) \mathcal{O}(\omega^2)
    + \mathcal{O}(\omega^4)
  \right]
}
\begin{align}
  \GHDLF,
  \label{eq:G_HD_LF}
\end{align}
which clearly shows the cancellation of DC-like noises under the condition \cref{eq:fine-tuning}.
In \cref{fig:noise}a, $G(\omega)$ for various protocols is plotted.
After the cancellation, the leading contribution comes from the $\mathcal{O}(\omega^4)$ terms as
\newcommand{\GHDFT}{%
  G(\omega) \to \frac{\omega^4 P(\omega)}{576n^4} \gamma_e^2 \tau_e^2 \left(
    3\tau^2 - 3\tau\tau_e - \tau_N^2 + \tau_e^2
  \right)^2
}
\begin{align}
  \GHDFT.
  ~~
  (\text{fine-tuned})
  \label{eq:G_HD_FT}
\end{align}
According to this expression, combined with the relationship $\tau=\tau_N+\tau_e+2n\tov$, effects of $\tov$ start to modify the noise amplitude by $\mathcal{O}(1)$ when the total operation time of the swap gates becomes comparable to the total sensing time, $2n\tov \gtrsim \tau_N$.
This is confirmed by numerical calculations shown in \cref{sec:noise}, and visualised in \cref{fig:noise}b.

To look at the noise from a sensing perspective, we perform simulations in the same way as for the coherence times (see \cref{sec:simulation} for details). In \cref{fig:noise}c, the effect of DC noise on the final result is plotted, while measuring a magnetic field gradient. Note that for NV centres, the gradient chosen for clear visualisation in this example would be the rather large $\SI{2e4}{T/m}$. The detectable gradient strongly depends on the volume with the same gradient and the distance between the spins used, as even a field difference of merely $\SI{1}{pT}$ gives a gradient of $\SI{6}{mT/m}$ for a single NV centre. Looking at pink noise in \cref{fig:noise}d, there is uncertainty in the final measurement. Increasing the repetitions $n$, this uncertainty decreases, as depicted in \cref{fig:noise}e.

\section*{Dark matter search}

The proposed hybrid-spin decoupling protocol is also suitable for exploring exotic spin-dependent interactions whose coupling strengths are irrelevant to the gyromagnetic ratios.
As an example, we focus on the ultralight axion dark matter searches with NV centres, which was originally considered in \cite{Chigusa:2023roq,Chigusa:2024psk}.
An ultralight axion dark matter with mass $m_a\ll \si{eV}$ is well described by the classical field
\newcommand{\atx}{%
  a(t, \bm{x}) = a_0 \sin(m_a t + m_a \bm{v}_a\cdot\bm{x} + \phi)
}
\begin{align}
  \atx,
  \label{eq:axion_field}
\end{align}
where $\bm{v}_a$ and $\phi$ are the axion velocity and the oscillation phase, both of which change randomly after the coherence time $\tau_a$.
As can be seen, the oscillation frequency is approximately identical to the axion mass $m_a$.
Through its interactions with fermions, the axion field results in the interaction Hamiltonian
\newcommand{\Hint}{%
  H_{\mathrm{int}}(t) = \sum_\chi \gamma_\chi \bm{B}_\chi(t) \cdot \bm{S}_\chi
}
\begin{align}
  \Hint,
  \label{eq:Hint}
\end{align}
where $\chi=e,p,n,\dots$ represents the fermion species, and $\gamma_\chi$ and $\bm{S}_\chi$ are the gyromagnetic ratio and the spin operator of $\chi$, respectively.
The effective magnetic field is defined as
\newcommand{\Beff}{%
  \gamma_\chi \bm{B}_\chi(t) = \frac{g_{a\chi\chi}}{m_\chi} \sqrt{2\rho_a} \bm{v}_a \cos(m_a t + \phi)
}
\begin{align}
  \Beff,
  \label{eq:B_eff_chi}
\end{align}
where $m_\chi$ is the mass of the fermion, and $\rho_{\mathrm{DM}} \sim \SI{0.45}{GeV}$ is the observed energy density of dark matter related to the axion oscillation amplitude $a_0$ as~\cref{eq:rho_DM}.
Importantly, these effective magnetic fields are determined by the $\chi$-dependent axion-fermion coupling constants, $g_{a\chi\chi}$.

\begin{figure}[!h]
  \centering
  \includegraphics{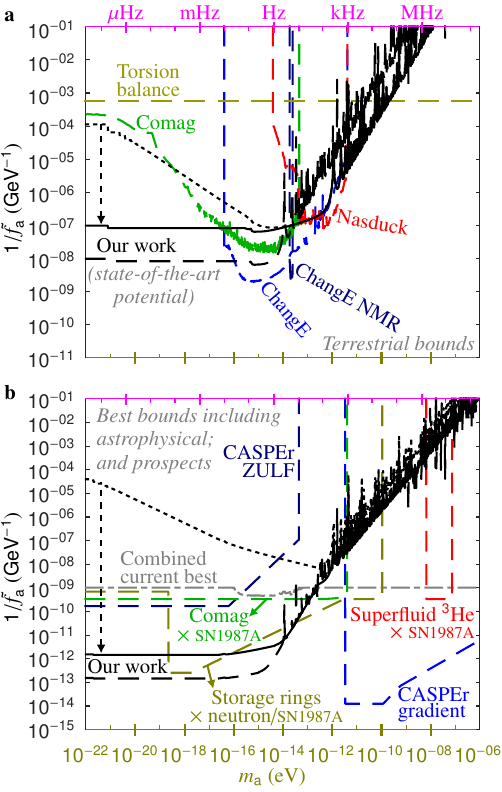}
  \caption{Prospected dark matter constraints. \textbf{a}. Constraints on the axion decay constant vs axion mass as set by various experiments on earth~\cite{Adelberger:2006dh,Bloch:2021vnn,Bloch:2022kjm,Xu:2023vfn,Wei:2023rzs,Gavilan-Martin:2024nlo} (coloured dashed lines~\cite{OHare:2020}), including recent work with comagnetometers (green dashed line) which includes over two months of collected data~\cite{Gavilan:2025}. Our potential constraints with a current state-of-the-art ensemble sample ($10^{12}$\,NVs) for a one-month long measurement is plotted in black. The dotted line indicates the Ramsey result from previous work~\cite{Chigusa:2024psk}, which is affected by noise. The solid line is for the presented work, which negates this magnetic noise. Finally, the dashed line is when utilising the potential increase in coherence time for large $n$ at low temperature. \textbf{b} The current best constraint based on neutron stars, K-$^3$He comagnetometers, ChangE and SN1987A~\cite{Lella:2023bfb} (grey dashed-dotted line) is compared with potential future experiments~\cite{JacksonKimball:2017elr,Wu:2019exd,Garcon:2019inh,Bloch:2019lcy,Gao:2022nuq,Brandenstein:2022eif} (coloured dashed lines~\cite{OHare:2020}). For our prospects, we look at a large sample ($10^{20}$\,NVs) for a year-long measurement. This illuminates that the resilience to noise becomes more important when the sensitivity becomes better.
  }
  \label{fig:darkmatter}
\end{figure}

Owing to the expected weak signal for axion dark matter searches, our approach is based on the repeated measurements and a successive statistical analysis to gain sensitivities.
Let $\Nobs$ be the number of repetitions, and $\phi_j$ be the $j$-th measurement result with $j=1,\dots,\Nobs$ labelling each measurement. 
Following the statistical treatment proposed in~\cite{Chigusa:2024psk}, we consider a power spectral density of the time sequence data
\begin{align}
  P_k \equiv \frac{\tau^2}{\tobs} \sum_{j,j'} e^{i\omega_k (t_j-t_{j'})} \phi_j \phi_{j'},
\end{align}
where $\tobs\equiv \Nobs\tau$ is the total experimental run time, $\omega_k = 2\pi k/\tobs$ is the discretised angular frequency, and $t_j\equiv j \tau$.
See \cref{sec:app_sensitivity} for the detailed calculation of the ensemble-averaged power spectral density, $\PSD{k}$, defined in \cref{eq:PSD_def}.
An important point is that, due to the Fourier transform in the above definition, $P_k$ is a single-mode observable focused on $\omega_k$, which is suitable for detecting a fixed frequency signal with $m_a \sim \omega_k$.
Moreover, the external magnetic noise contribution to $\PSD{k}$ turns out to be $G(\omega_k)$ (see \cref{eq:BPSD} and below), which implies that the noise suppression mechanism described above is indeed beneficial for this approach when $\omega_k \lesssim \SI{1}{kHz}$.

In \cref{fig:darkmatter}a, we compare the potential of our protocol with current terrestrial bounds. The sensitivities are shown in terms of the combined coupling 
\begin{align}
  \tilde{f}_a^{-1} \equiv \frac{1}{2} \left| \frac{g_{app}}{m_p} + \frac{g_{ann}}{m_n} \right|,  
\end{align}
where $g_{app}$ and $g_{ann}$ are axion couplings with protons and neutrons, respectively (see \cref{sec:app_sensitivity} for detailed definitions). Here, we assume a diamond ensemble sample with $10^{12}$\,NVs (close to the current state-of-the-art sample~\cite{Barry:2024}), for a one-month measurement. Merely using the nuclear spin~\cite{Chigusa:2023roq} is susceptible to the assumed noise \cref{eq:P_omega_assumed}, as visible by the black dotted line.
However, with our hybrid-spin decoupling protocol, the noise is cancelled, and a broadband search is possible (black line for single repetition $n=1$), similar to comagnetometer results~\cite{Gavilan:2025}.
With increased $n$, further improvement is possible. Note that compared to low-temperature improvements of $T_2$~\cite{BarGill:2013} which is about three orders of magnitude at $77$~K, we indicate results for two orders only. The reason is that the required swap gates have limited fidelity, although there is much room for improvement with decoherence-protected swap gates~\cite{Bartling:2025}.

To compare to prospects of various proposals, we plotted these in \cref{fig:darkmatter}b. For our proposal, we look at a diamond ensemble sample with $10^{20}$\,NVs for a one-year measurement. Although this is currently highly challenging, we want to emphasise the importance of the hybrid-spin decoupling protocol for increased sensitivities; a much smaller ensemble of about $10^{14}$\,NVs, even without the improvement of coherence, would be able to surpass the current best bounds given by astrophysical observations. As visible from the black dotted line in \cref{fig:darkmatter}b, with improved sensitivity, the noise cancellation becomes more important, showing over seven orders of magnitude difference for the lowest masses. Note that if such a level is reached, other error sources, such as systematic errors, become more important. For example, the accuracy of the gyromagnetic ratios and the precision of the delays play an important role as well to gain the improved sensitivity. Adding shielding to alleviate the amount of noise required to cancel could help in this scenario.

\section*{Future directions}\label{sec:discussion}

We presented a hybrid-spin decoupling protocol that is resistant to noise. It uniquely allows to measure certain DC fields, while cancelling AC noise, thus allowing for long coherence times and high sensitivities. The requirement for the DC field to be measured is that it must interact effectively differently with each spin compared to magnetic noise. An example is to measure a gradient field, thus when there is a difference in the magnetic field at the locations of the two spins, as is simulated in \cref{fig:noise}.
Another example is a single spin measurement, where sensor spins placed near the target spin experience different magnetic fields through dipole–dipole interactions, the differences arising from their spatial separation.
For such applications, the spins could be of the same type, although an additional $\pi$-pulse after the swap gate would be required in the protocol. For example, using coupled NV centres about $\SI{10}{nm}$ apart~\cite{haruyama:2019} would greatly enhance the sensitivity to a gradient.
Therefore, the applicability of this protocol extends beyond NV centres and can be broadly utilised in various two-spin-based sensing platforms.

The gain is most substantial for new-physics searches. For example, when looking at comagnetometry setups for dark matter searches, noise resilience plays an important role to achieve the required sensitivity to a faint signal.
For our method, the noise resistance is achieved in a completely different way. It decouples from noise, and unlike conventional decoupling sequences, broadband DC sensing remains possible, which is crucial for dark matter searches to cover a broad mass range. It allows for reducing noise with orders of magnitude higher frequencies ($\sim1$ Hz for comagnetometry~\cite{Kornack:2002,Klinger:2023} vs $\sim1$ kHz and higher for larger $n$), and further improvement is possible at low temperatures, options not available for comagnetometers.
Moreover, other solid-state systems with suitable spin combinations, such as silicon carbide or hexagonal boron nitride, could apply this method as well. For future work, we plan to implement this protocol with NV centres in diamond, initially with the aim of improving the current best terrestrial bounds on axion dark matter in the low mass regime~\cite{Gavilan:2025}.

\backmatter



\bmhead{Acknowledgements}

SC thanks Itay M. Bloch for fruitful discussions.
This work is supported by the Simons Foundation.
This work is supported by the U.S. Department of Energy, Office of Science, the BNL C2QA award under grant Contract Number DESC0012704 (SUBK\#390034).
This work is supported by JSPS KAKENHI (Grant Number 24K07010 [KN]).
This work is supported by World Premier International Research Center Initiative (WPI), MEXT, Japan.
This work is supported by JST Moonshot (Grant Number JPMJMS226C), CREST (JPMJCR23I5), and Presto JST (JPMJPR245B).





\bibliography{sn-bibliography}

\begin{thebibliography}{10}
\expandafter\ifx\csname url\endcsname\relax
  \def\url#1{\burl{#1}}\fi
\expandafter\ifx\csname urlprefix\endcsname\relax\def\urlprefix{URL }\fi
\providecommand{\bibinfo}[2]{#2}
\providecommand{\eprint}[2][]{\url{#2}}
\providecommand{\doi}[1]{\url{https://doi.org/#1}}
\bibcommenthead

\bibitem{Degen:2017}
\bibinfo{author}{Degen, C.~L.}, \bibinfo{author}{Reinhard, F.} \& \bibinfo{author}{Cappellaro, P.}
\newblock \bibinfo{title}{Quantum sensing}.
\newblock \emph{\bibinfo{journal}{Rev. Mod. Phys.}} \textbf{\bibinfo{volume}{89}}, \bibinfo{pages}{035002} (\bibinfo{year}{2017}).
\newblock \urlprefix\url{https://link.aps.org/doi/10.1103/RevModPhys.89.035002}.

\bibitem{Herbschleb:2019}
\bibinfo{author}{Herbschleb, E.} \emph{et~al.}
\newblock \bibinfo{title}{Ultra-long coherence times amongst room-temperature solid-state spins}.
\newblock \emph{\bibinfo{journal}{Nature communications}} \textbf{\bibinfo{volume}{10}}, \bibinfo{pages}{3766} (\bibinfo{year}{2019}).

\bibitem{Kenny:2025}
\bibinfo{author}{Kenny, J.} \emph{et~al.}
\newblock \bibinfo{title}{Quantum sensing enhancement through a nuclear spin register in nitrogen-vacancy centers in diamond}.
\newblock \emph{\bibinfo{journal}{Applied Physics Reviews}} \textbf{\bibinfo{volume}{12}}, \bibinfo{pages}{021323} (\bibinfo{year}{2025}).
\newblock \urlprefix\url{https://doi.org/10.1063/5.0235057}.

\bibitem{Wang:2017}
\bibinfo{author}{Wang, Z.-Y.}, \bibinfo{author}{Casanova, J.} \& \bibinfo{author}{Plenio, M.~B.}
\newblock \bibinfo{title}{Delayed entanglement echo for individual control of a large number of nuclear spins}.
\newblock \emph{\bibinfo{journal}{Nature communications}} \textbf{\bibinfo{volume}{8}}, \bibinfo{pages}{14660} (\bibinfo{year}{2017}).

\bibitem{Ajoy:2015}
\bibinfo{author}{Ajoy, A.}, \bibinfo{author}{Bissbort, U.}, \bibinfo{author}{Lukin, M.~D.}, \bibinfo{author}{Walsworth, R.~L.} \& \bibinfo{author}{Cappellaro, P.}
\newblock \bibinfo{title}{Atomic-scale nuclear spin imaging using quantum-assisted sensors in diamond}.
\newblock \emph{\bibinfo{journal}{Phys. Rev. X}} \textbf{\bibinfo{volume}{5}}, \bibinfo{pages}{011001} (\bibinfo{year}{2015}).
\newblock \urlprefix\url{https://link.aps.org/doi/10.1103/PhysRevX.5.011001}.

\bibitem{Chen:2023}
\bibinfo{author}{Chen, X.-Y.} \emph{et~al.}
\newblock \bibinfo{title}{Extending dephasing time of nitrogen-vacancy center in diamond by suppressing nuclear spin noise}.
\newblock \emph{\bibinfo{journal}{Phys. Rev. B}} \textbf{\bibinfo{volume}{108}}, \bibinfo{pages}{174111} (\bibinfo{year}{2023}).
\newblock \urlprefix\url{https://link.aps.org/doi/10.1103/PhysRevB.108.174111}.

\bibitem{Dreau:2013}
\bibinfo{author}{Dr\'eau, A.}, \bibinfo{author}{Spinicelli, P.}, \bibinfo{author}{Maze, J.~R.}, \bibinfo{author}{Roch, J.-F.} \& \bibinfo{author}{Jacques, V.}
\newblock \bibinfo{title}{Single-shot readout of multiple nuclear spin qubits in diamond under ambient conditions}.
\newblock \emph{\bibinfo{journal}{Phys. Rev. Lett.}} \textbf{\bibinfo{volume}{110}}, \bibinfo{pages}{060502} (\bibinfo{year}{2013}).
\newblock \urlprefix\url{https://link.aps.org/doi/10.1103/PhysRevLett.110.060502}.

\bibitem{Xie:2021}
\bibinfo{author}{Xie, T.} \emph{et~al.}
\newblock \bibinfo{title}{Beating the standard quantum limit under ambient conditions with solid-state spins}.
\newblock \emph{\bibinfo{journal}{Science Advances}} \textbf{\bibinfo{volume}{7}}, \bibinfo{pages}{eabg9204} (\bibinfo{year}{2021}).
\newblock \urlprefix\url{https://www.science.org/doi/abs/10.1126/sciadv.abg9204}.

\bibitem{Burgler:2023}
\bibinfo{author}{B\"{u}rgler, B.} \emph{et~al.}
\newblock \bibinfo{title}{All-optical nuclear quantum sensing using nitrogen-vacancy centers in diamond}.
\newblock \emph{\bibinfo{journal}{npj Quantum Information}} \textbf{\bibinfo{volume}{9}}, \bibinfo{pages}{56} (\bibinfo{year}{2023}).

\bibitem{Kornack:2002}
\bibinfo{author}{Kornack, T.~W.} \& \bibinfo{author}{Romalis, M.~V.}
\newblock \bibinfo{title}{Dynamics of two overlapping spin ensembles interacting by spin exchange}.
\newblock \emph{\bibinfo{journal}{Phys. Rev. Lett.}} \textbf{\bibinfo{volume}{89}}, \bibinfo{pages}{253002} (\bibinfo{year}{2002}).
\newblock \urlprefix\url{https://link.aps.org/doi/10.1103/PhysRevLett.89.253002}.

\bibitem{Klinger:2023}
\bibinfo{author}{Klinger, E.} \emph{et~al.}
\newblock \bibinfo{title}{Optimization of nuclear polarization in an alkali-noble gas comagnetometer}.
\newblock \emph{\bibinfo{journal}{Phys. Rev. Appl.}} \textbf{\bibinfo{volume}{19}}, \bibinfo{pages}{044092} (\bibinfo{year}{2023}).
\newblock \urlprefix\url{https://link.aps.org/doi/10.1103/PhysRevApplied.19.044092}.

\bibitem{Padniuk:2024}
\bibinfo{author}{Padniuk, M.} \emph{et~al.}
\newblock \bibinfo{title}{Universal determination of comagnetometer response to spin couplings}.
\newblock \emph{\bibinfo{journal}{Phys. Rev. Res.}} \textbf{\bibinfo{volume}{6}}, \bibinfo{pages}{013339} (\bibinfo{year}{2024}).
\newblock \urlprefix\url{https://link.aps.org/doi/10.1103/PhysRevResearch.6.013339}.

\bibitem{Gavilan:2025}
\bibinfo{author}{Gavilan-Martin, D.} \emph{et~al.}
\newblock \bibinfo{title}{Searching for dark matter with a spin-based interferometer}.
\newblock \emph{\bibinfo{journal}{Nature communications}} \textbf{\bibinfo{volume}{16}}, \bibinfo{pages}{4953} (\bibinfo{year}{2025}).

\bibitem{MeiboomGill:1958}
\bibinfo{author}{Meiboom, S.} \& \bibinfo{author}{Gill, D.}
\newblock \bibinfo{title}{Modified spin‐echo method for measuring nuclear relaxation times}.
\newblock \emph{\bibinfo{journal}{Review of Scientific Instruments}} \textbf{\bibinfo{volume}{29}}, \bibinfo{pages}{688--691} (\bibinfo{year}{1958}).
\newblock \urlprefix\url{https://doi.org/10.1063/1.1716296}.

\bibitem{BarGill:2013}
\bibinfo{author}{Bar-Gill, N.}, \bibinfo{author}{Pham, L.}, \bibinfo{author}{Jarmola, A.}, \bibinfo{author}{Budker, D.} \& \bibinfo{author}{Walsworth, R.}
\newblock \bibinfo{title}{Solid-state electronic spin coherence time approaching one second}.
\newblock \emph{\bibinfo{journal}{Nature communications}} \textbf{\bibinfo{volume}{4}}, \bibinfo{pages}{1743} (\bibinfo{year}{2013}).

\bibitem{Maurer:2012}
\bibinfo{author}{Maurer, P.~C.} \emph{et~al.}
\newblock \bibinfo{title}{Room-temperature quantum bit memory exceeding one second}.
\newblock \emph{\bibinfo{journal}{Science}} \textbf{\bibinfo{volume}{336}}, \bibinfo{pages}{1283--1286} (\bibinfo{year}{2012}).
\newblock \urlprefix\url{https://www.science.org/doi/abs/10.1126/science.1220513}.

\bibitem{Ilias:2024}
\bibinfo{author}{Ilias, T.}, \bibinfo{author}{Yang, D.}, \bibinfo{author}{Huelga, S.~F.} \& \bibinfo{author}{Plenio, M.~B.}
\newblock \bibinfo{title}{Criticality-enhanced electric field gradient sensor with single trapped ions}.
\newblock \emph{\bibinfo{journal}{npj Quantum Information}} \textbf{\bibinfo{volume}{10}}, \bibinfo{pages}{36} (\bibinfo{year}{2024}).

\bibitem{Kornack:2005}
\bibinfo{author}{Kornack, T.~W.}, \bibinfo{author}{Ghosh, R.~K.} \& \bibinfo{author}{Romalis, M.~V.}
\newblock \bibinfo{title}{Nuclear spin gyroscope based on an atomic comagnetometer}.
\newblock \emph{\bibinfo{journal}{Phys. Rev. Lett.}} \textbf{\bibinfo{volume}{95}}, \bibinfo{pages}{230801} (\bibinfo{year}{2005}).
\newblock \urlprefix\url{https://link.aps.org/doi/10.1103/PhysRevLett.95.230801}.

\bibitem{Fang:2012}
\bibinfo{author}{Fang, J.} \& \bibinfo{author}{Qin, J.}
\newblock \bibinfo{title}{Advances in atomic gyroscopes: A view from inertial navigation applications}.
\newblock \emph{\bibinfo{journal}{Sensors}} \textbf{\bibinfo{volume}{12}}, \bibinfo{pages}{6331--6346} (\bibinfo{year}{2012}).
\newblock \urlprefix\url{https://www.mdpi.com/1424-8220/12/5/6331}.

\bibitem{PhysRevLett.124.081803}
\bibinfo{author}{Abel, C.} \emph{et~al.}
\newblock \bibinfo{title}{Measurement of the permanent electric dipole moment of the neutron}.
\newblock \emph{\bibinfo{journal}{Phys. Rev. Lett.}} \textbf{\bibinfo{volume}{124}}, \bibinfo{pages}{081803} (\bibinfo{year}{2020}).
\newblock \urlprefix\url{https://link.aps.org/doi/10.1103/PhysRevLett.124.081803}.

\bibitem{PhysRevLett.123.143003}
\bibinfo{author}{Sachdeva, N.} \emph{et~al.}
\newblock \bibinfo{title}{New limit on the permanent electric dipole moment of $^{129}\mathrm{Xe}$ using $^{3}\mathrm{He}$ comagnetometry and squid detection}.
\newblock \emph{\bibinfo{journal}{Phys. Rev. Lett.}} \textbf{\bibinfo{volume}{123}}, \bibinfo{pages}{143003} (\bibinfo{year}{2019}).
\newblock \urlprefix\url{https://link.aps.org/doi/10.1103/PhysRevLett.123.143003}.

\bibitem{PhysRevA.100.022505}
\bibinfo{author}{Allmendinger, F.} \emph{et~al.}
\newblock \bibinfo{title}{Measurement of the permanent electric dipole moment of the $^{129}\mathrm{Xe}$ atom}.
\newblock \emph{\bibinfo{journal}{Phys. Rev. A}} \textbf{\bibinfo{volume}{100}}, \bibinfo{pages}{022505} (\bibinfo{year}{2019}).
\newblock \urlprefix\url{https://link.aps.org/doi/10.1103/PhysRevA.100.022505}.

\bibitem{Vasilakis:2008yn}
\bibinfo{author}{Vasilakis, G.}, \bibinfo{author}{Brown, J.~M.}, \bibinfo{author}{Kornack, T.~W.} \& \bibinfo{author}{Romalis, M.~V.}
\newblock \bibinfo{title}{{Limits on new long range nuclear spin-dependent forces set with a K - He-3 co-magnetometer}}.
\newblock \emph{\bibinfo{journal}{Phys. Rev. Lett.}} \textbf{\bibinfo{volume}{103}}, \bibinfo{pages}{261801} (\bibinfo{year}{2009}).

\bibitem{Lee:2018}
\bibinfo{author}{Lee, J.}, \bibinfo{author}{Almasi, A.} \& \bibinfo{author}{Romalis, M.}
\newblock \bibinfo{title}{Improved limits on spin-mass interactions}.
\newblock \emph{\bibinfo{journal}{Phys. Rev. Lett.}} \textbf{\bibinfo{volume}{120}}, \bibinfo{pages}{161801} (\bibinfo{year}{2018}).
\newblock \urlprefix\url{https://link.aps.org/doi/10.1103/PhysRevLett.120.161801}.

\bibitem{Almasi:2018cob}
\bibinfo{author}{Almasi, A.}, \bibinfo{author}{Lee, J.}, \bibinfo{author}{Winarto, H.}, \bibinfo{author}{Smiciklas, M.} \& \bibinfo{author}{Romalis, M.~V.}
\newblock \bibinfo{title}{{New Limits on Anomalous Spin-Spin Interactions}}.
\newblock \emph{\bibinfo{journal}{Phys. Rev. Lett.}} \textbf{\bibinfo{volume}{125}}, \bibinfo{pages}{201802} (\bibinfo{year}{2020}).

\bibitem{Zhang:2023}
\bibinfo{author}{Zhang, S.-B.} \emph{et~al.}
\newblock \bibinfo{title}{Search for spin-dependent gravitational interactions at earth range}.
\newblock \emph{\bibinfo{journal}{Phys. Rev. Lett.}} \textbf{\bibinfo{volume}{130}}, \bibinfo{pages}{201401} (\bibinfo{year}{2023}).
\newblock \urlprefix\url{https://link.aps.org/doi/10.1103/PhysRevLett.130.201401}.

\bibitem{Kornack:thesis}
\bibinfo{author}{Kornack, T.~W.}
\newblock \emph{\bibinfo{title}{A test of CPT and Lorentz symmetry using a potassium-helium-3 co-magnetometer}}.
\newblock Ph.D. thesis (\bibinfo{year}{2005}).
\newblock \urlprefix\url{https://www.proquest.com/dissertations-theses/test-cpt-lorentz-symmetry-using-potassium-helium/docview/305415550/se-2}.
\newblock \bibinfo{note}{Copyright - Database copyright ProQuest LLC; ProQuest does not claim copyright in the individual underlying works; Last updated - 2023-03-03}.

\bibitem{Brown:2010dt}
\bibinfo{author}{Brown, J.~M.}, \bibinfo{author}{Smullin, S.~J.}, \bibinfo{author}{Kornack, T.~W.} \& \bibinfo{author}{Romalis, M.~V.}
\newblock \bibinfo{title}{{New limit on Lorentz and CPT-violating neutron spin interactions}}.
\newblock \emph{\bibinfo{journal}{Phys. Rev. Lett.}} \textbf{\bibinfo{volume}{105}}, \bibinfo{pages}{151604} (\bibinfo{year}{2010}).

\bibitem{Smiciklas:2011}
\bibinfo{author}{Smiciklas, M.}, \bibinfo{author}{Brown, J.~M.}, \bibinfo{author}{Cheuk, L.~W.}, \bibinfo{author}{Smullin, S.~J.} \& \bibinfo{author}{Romalis, M.~V.}
\newblock \bibinfo{title}{New test of local lorentz invariance using a $^{21}\mathrm{Ne}\mathrm{\text{\ensuremath{-}}}\mathrm{Rb}\mathrm{\text{\ensuremath{-}}}\mathbf{K}$ comagnetometer}.
\newblock \emph{\bibinfo{journal}{Phys. Rev. Lett.}} \textbf{\bibinfo{volume}{107}}, \bibinfo{pages}{171604} (\bibinfo{year}{2011}).
\newblock \urlprefix\url{https://link.aps.org/doi/10.1103/PhysRevLett.107.171604}.

\bibitem{degen:2008}
\bibinfo{author}{Degen, C.}
\newblock \bibinfo{title}{Scanning magnetic field microscope with a diamond single-spin sensor}.
\newblock \emph{\bibinfo{journal}{Applied Physics Letters}} \textbf{\bibinfo{volume}{92}} (\bibinfo{year}{2008}).

\bibitem{Neumann:2010}
\bibinfo{author}{Neumann, P.} \emph{et~al.}
\newblock \bibinfo{title}{Single-shot readout of a single nuclear spin}.
\newblock \emph{\bibinfo{journal}{Science}} \textbf{\bibinfo{volume}{329}}, \bibinfo{pages}{542--544} (\bibinfo{year}{2010}).
\newblock \urlprefix\url{https://www.science.org/doi/abs/10.1126/science.1189075}.

\bibitem{PhysRev.125.912}
\bibinfo{author}{Klauder, J.~R.} \& \bibinfo{author}{Anderson, P.~W.}
\newblock \bibinfo{title}{Spectral diffusion decay in spin resonance experiments}.
\newblock \emph{\bibinfo{journal}{Phys. Rev.}} \textbf{\bibinfo{volume}{125}}, \bibinfo{pages}{912--932} (\bibinfo{year}{1962}).
\newblock \urlprefix\url{https://link.aps.org/doi/10.1103/PhysRev.125.912}.

\bibitem{PhysRevLett.97.087601}
\bibinfo{author}{Hanson, R.}, \bibinfo{author}{Mendoza, F.~M.}, \bibinfo{author}{Epstein, R.~J.} \& \bibinfo{author}{Awschalom, D.~D.}
\newblock \bibinfo{title}{Polarization and readout of coupled single spins in diamond}.
\newblock \emph{\bibinfo{journal}{Phys. Rev. Lett.}} \textbf{\bibinfo{volume}{97}}, \bibinfo{pages}{087601} (\bibinfo{year}{2006}).
\newblock \urlprefix\url{https://link.aps.org/doi/10.1103/PhysRevLett.97.087601}.

\bibitem{PhysRevB.87.115122}
\bibinfo{author}{Wang, Z.-H.} \& \bibinfo{author}{Takahashi, S.}
\newblock \bibinfo{title}{Spin decoherence and electron spin bath noise of a nitrogen-vacancy center in diamond}.
\newblock \emph{\bibinfo{journal}{Phys. Rev. B}} \textbf{\bibinfo{volume}{87}}, \bibinfo{pages}{115122} (\bibinfo{year}{2013}).
\newblock \urlprefix\url{https://link.aps.org/doi/10.1103/PhysRevB.87.115122}.

\bibitem{doi:10.7566/JPSJ.89.054708}
\bibinfo{author}{Hayashi, K.} \emph{et~al.}
\newblock \bibinfo{title}{Experimental and theoretical analysis of noise strength and environmental correlation time for ensembles of nitrogen-vacancy centers in diamond}.
\newblock \emph{\bibinfo{journal}{Journal of the Physical Society of Japan}} \textbf{\bibinfo{volume}{89}}, \bibinfo{pages}{054708} (\bibinfo{year}{2020}).
\newblock \urlprefix\url{https://doi.org/10.7566/JPSJ.89.054708}.

\bibitem{averin2016suppression}
\bibinfo{author}{Averin, D.} \emph{et~al.}
\newblock \bibinfo{title}{Suppression of dephasing by qubit motion in superconducting circuits}.
\newblock \emph{\bibinfo{journal}{Physical review letters}} \textbf{\bibinfo{volume}{116}}, \bibinfo{pages}{010501} (\bibinfo{year}{2016}).

\bibitem{matsuzaki2016hybrid}
\bibinfo{author}{Matsuzaki, Y.} \emph{et~al.}
\newblock \bibinfo{title}{Hybrid quantum magnetic-field sensor with an electron spin and a nuclear spin in diamond}.
\newblock \emph{\bibinfo{journal}{Physical Review A}} \textbf{\bibinfo{volume}{94}}, \bibinfo{pages}{052330} (\bibinfo{year}{2016}).

\bibitem{Barry:2020}
\bibinfo{author}{Barry, J.~F.} \emph{et~al.}
\newblock \bibinfo{title}{Sensitivity optimization for nv-diamond magnetometry}.
\newblock \emph{\bibinfo{journal}{Rev. Mod. Phys.}} \textbf{\bibinfo{volume}{92}}, \bibinfo{pages}{015004} (\bibinfo{year}{2020}).
\newblock \urlprefix\url{https://link.aps.org/doi/10.1103/RevModPhys.92.015004}.

\bibitem{Mizuochi:2009}
\bibinfo{author}{Mizuochi, N.} \emph{et~al.}
\newblock \bibinfo{title}{Coherence of single spins coupled to a nuclear spin bath of varying density}.
\newblock \emph{\bibinfo{journal}{Physical Review B}} \textbf{\bibinfo{volume}{80}}, \bibinfo{pages}{041201 (R)} (\bibinfo{year}{2009}).
\newblock \urlprefix\url{https://doi.org/10.1103/PhysRevB.80.041201}.

\bibitem{Wolf:2015}
\bibinfo{author}{Wolf, T.} \emph{et~al.}
\newblock \bibinfo{title}{Subpicotesla diamond magnetometry}.
\newblock \emph{\bibinfo{journal}{Phys. Rev. X}} \textbf{\bibinfo{volume}{5}}, \bibinfo{pages}{041001} (\bibinfo{year}{2015}).
\newblock \urlprefix\url{https://link.aps.org/doi/10.1103/PhysRevX.5.041001}.

\bibitem{Chigusa:2023roq}
\bibinfo{author}{Chigusa, S.}, \bibinfo{author}{Hazumi, M.}, \bibinfo{author}{Herbschleb, E.~D.}, \bibinfo{author}{Mizuochi, N.} \& \bibinfo{author}{Nakayama, K.}
\newblock \bibinfo{title}{{Light dark matter search with nitrogen-vacancy centers in diamonds}}.
\newblock \emph{\bibinfo{journal}{JHEP}} \textbf{\bibinfo{volume}{03}}, \bibinfo{pages}{083} (\bibinfo{year}{2025}).

\bibitem{Chigusa:2024psk}
\bibinfo{author}{Chigusa, S.} \emph{et~al.}
\newblock \bibinfo{title}{{Nuclear spin metrology with nitrogen vacancy center in diamond for axion dark matter detection}}.
\newblock \emph{\bibinfo{journal}{Phys. Rev. D}} \textbf{\bibinfo{volume}{111}}, \bibinfo{pages}{075028} (\bibinfo{year}{2025}).

\bibitem{Adelberger:2006dh}
\bibinfo{author}{Adelberger, E.~G.} \emph{et~al.}
\newblock \bibinfo{title}{{Particle Physics Implications of a Recent Test of the Gravitational Inverse Sqaure Law}}.
\newblock \emph{\bibinfo{journal}{Phys. Rev. Lett.}} \textbf{\bibinfo{volume}{98}}, \bibinfo{pages}{131104} (\bibinfo{year}{2007}).

\bibitem{Bloch:2021vnn}
\bibinfo{author}{Bloch, I.~M.} \emph{et~al.}
\newblock \bibinfo{title}{{New constraints on axion-like dark matter using a Floquet quantum detector}}.
\newblock \emph{\bibinfo{journal}{Sci. Adv.}} \textbf{\bibinfo{volume}{8}}, \bibinfo{pages}{abl8919} (\bibinfo{year}{2022}).

\bibitem{Bloch:2022kjm}
\bibinfo{author}{Bloch, I.~M.} \emph{et~al.}
\newblock \bibinfo{title}{{Constraints on axion-like dark matter from a SERF comagnetometer}}.
\newblock \emph{\bibinfo{journal}{Nature Commun.}} \textbf{\bibinfo{volume}{14}}, \bibinfo{pages}{5784} (\bibinfo{year}{2023}).

\bibitem{Xu:2023vfn}
\bibinfo{author}{Xu, Z.} \emph{et~al.}
\newblock \bibinfo{title}{{Constraining ultralight dark matter through an accelerated resonant search}}.
\newblock \emph{\bibinfo{journal}{Commun. Phys.}} \textbf{\bibinfo{volume}{7}}, \bibinfo{pages}{226} (\bibinfo{year}{2024}).

\bibitem{Wei:2023rzs}
\bibinfo{author}{Wei, K.} \emph{et~al.}
\newblock \bibinfo{title}{{Dark matter search with a resonantly-coupled hybrid spin system}}.
\newblock \emph{\bibinfo{journal}{Rept. Prog. Phys.}} \textbf{\bibinfo{volume}{88}}, \bibinfo{pages}{057801} (\bibinfo{year}{2025}).

\bibitem{Gavilan-Martin:2024nlo}
\bibinfo{author}{Gavilan-Martin, D.} \emph{et~al.}
\newblock \bibinfo{title}{{Searching for dark matter with a spin-based interferometer}}.
\newblock \emph{\bibinfo{journal}{Nature Commun.}} \textbf{\bibinfo{volume}{16}}, \bibinfo{pages}{4953} (\bibinfo{year}{2025}).

\bibitem{OHare:2020}
\bibinfo{author}{O'Hare, C.}
\newblock \bibinfo{title}{cajohare/axionlimits: Axionlimits} (\bibinfo{year}{2020}).
\newblock \urlprefix\url{https://doi.org/10.5281/zenodo.3932430}.

\bibitem{Lella:2023bfb}
\bibinfo{author}{Lella, A.} \emph{et~al.}
\newblock \bibinfo{title}{{Getting the most on supernova axions}}.
\newblock \emph{\bibinfo{journal}{Phys. Rev. D}} \textbf{\bibinfo{volume}{109}}, \bibinfo{pages}{023001} (\bibinfo{year}{2024}).

\bibitem{JacksonKimball:2017elr}
\bibinfo{author}{Jackson~Kimball, D.~F.} \emph{et~al.}
\newblock \bibinfo{title}{{Overview of the Cosmic Axion Spin Precession Experiment (CASPEr)}}.
\newblock \emph{\bibinfo{journal}{Springer Proc. Phys.}} \textbf{\bibinfo{volume}{245}}, \bibinfo{pages}{105--121} (\bibinfo{year}{2020}).

\bibitem{Wu:2019exd}
\bibinfo{author}{Wu, T.} \emph{et~al.}
\newblock \bibinfo{title}{{Search for Axionlike Dark Matter with a Liquid-State Nuclear Spin Comagnetometer}}.
\newblock \emph{\bibinfo{journal}{Phys. Rev. Lett.}} \textbf{\bibinfo{volume}{122}}, \bibinfo{pages}{191302} (\bibinfo{year}{2019}).

\bibitem{Garcon:2019inh}
\bibinfo{author}{Garcon, A.} \emph{et~al.}
\newblock \bibinfo{title}{{Constraints on bosonic dark matter from ultralow-field nuclear magnetic resonance}}.
\newblock \emph{\bibinfo{journal}{Sci. Adv.}} \textbf{\bibinfo{volume}{5}}, \bibinfo{pages}{eaax4539} (\bibinfo{year}{2019}).

\bibitem{Bloch:2019lcy}
\bibinfo{author}{Bloch, I.~M.}, \bibinfo{author}{Hochberg, Y.}, \bibinfo{author}{Kuflik, E.} \& \bibinfo{author}{Volansky, T.}
\newblock \bibinfo{title}{{Axion-like Relics: New Constraints from Old Comagnetometer Data}}.
\newblock \emph{\bibinfo{journal}{JHEP}} \textbf{\bibinfo{volume}{01}}, \bibinfo{pages}{167} (\bibinfo{year}{2020}).

\bibitem{Gao:2022nuq}
\bibinfo{author}{Gao, C.} \emph{et~al.}
\newblock \bibinfo{title}{{Axion Wind Detection with the Homogeneous Precession Domain of Superfluid Helium-3}}.
\newblock \emph{\bibinfo{journal}{Phys. Rev. Lett.}} \textbf{\bibinfo{volume}{129}}, \bibinfo{pages}{211801} (\bibinfo{year}{2022}).

\bibitem{Brandenstein:2022eif}
\bibinfo{author}{Brandenstein, C.} \emph{et~al.}
\newblock \bibinfo{title}{{Towards an electrostatic storage ring for fundamental physics measurements}}.
\newblock \emph{\bibinfo{journal}{EPJ Web Conf.}} \textbf{\bibinfo{volume}{282}}, \bibinfo{pages}{01017} (\bibinfo{year}{2023}).

\bibitem{Barry:2024}
\bibinfo{author}{Barry, J.~F.} \emph{et~al.}
\newblock \bibinfo{title}{Sensitive ac and dc magnetometry with nitrogen-vacancy-center ensembles in diamond}.
\newblock \emph{\bibinfo{journal}{Phys. Rev. Appl.}} \textbf{\bibinfo{volume}{22}}, \bibinfo{pages}{044069} (\bibinfo{year}{2024}).
\newblock \urlprefix\url{https://link.aps.org/doi/10.1103/PhysRevApplied.22.044069}.

\bibitem{Bartling:2025}
\bibinfo{author}{Bartling, H.} \emph{et~al.}
\newblock \bibinfo{title}{Universal high-fidelity quantum gates for spin qubits in diamond}.
\newblock \emph{\bibinfo{journal}{Phys. Rev. Appl.}} \textbf{\bibinfo{volume}{23}}, \bibinfo{pages}{034052} (\bibinfo{year}{2025}).
\newblock \urlprefix\url{https://link.aps.org/doi/10.1103/PhysRevApplied.23.034052}.

\bibitem{haruyama:2019}
\bibinfo{author}{Haruyama, M.} \emph{et~al.}
\newblock \bibinfo{title}{Triple nitrogen-vacancy centre fabrication by c5n4h n ion implantation}.
\newblock \emph{\bibinfo{journal}{Nature communications}} \textbf{\bibinfo{volume}{10}}, \bibinfo{pages}{2664} (\bibinfo{year}{2019}).

\bibitem{Waldherr2012}
\bibinfo{author}{Waldherr, G.} \emph{et~al.}
\newblock \bibinfo{title}{High-dynamic-range magnetometry with a single nuclear spin in diamond}.
\newblock \emph{\bibinfo{journal}{Nature Nanotechnology}} \textbf{\bibinfo{volume}{7}}, \bibinfo{pages}{105--108} (\bibinfo{year}{2012}).

\bibitem{Peccei:1977hh}
\bibinfo{author}{Peccei, R.~D.} \& \bibinfo{author}{Quinn, H.~R.}
\newblock \bibinfo{title}{{CP Conservation in the Presence of Instantons}}.
\newblock \emph{\bibinfo{journal}{Phys. Rev. Lett.}} \textbf{\bibinfo{volume}{38}}, \bibinfo{pages}{1440--1443} (\bibinfo{year}{1977}).

\bibitem{Weinberg:1977ma}
\bibinfo{author}{Weinberg, S.}
\newblock \bibinfo{title}{{A New Light Boson?}}
\newblock \emph{\bibinfo{journal}{Phys. Rev. Lett.}} \textbf{\bibinfo{volume}{40}}, \bibinfo{pages}{223--226} (\bibinfo{year}{1978}).

\bibitem{Wilczek:1977pj}
\bibinfo{author}{Wilczek, F.}
\newblock \bibinfo{title}{{Problem of Strong $P$ and $T$ Invariance in the Presence of Instantons}}.
\newblock \emph{\bibinfo{journal}{Phys. Rev. Lett.}} \textbf{\bibinfo{volume}{40}}, \bibinfo{pages}{279--282} (\bibinfo{year}{1978}).

\bibitem{PhysRevLett.43.103}
\bibinfo{author}{Kim, J.~E.}
\newblock \bibinfo{title}{Weak-interaction singlet and strong $\mathrm{CP}$ invariance}.
\newblock \emph{\bibinfo{journal}{Phys. Rev. Lett.}} \textbf{\bibinfo{volume}{43}}, \bibinfo{pages}{103--107} (\bibinfo{year}{1979}).
\newblock \urlprefix\url{https://link.aps.org/doi/10.1103/PhysRevLett.43.103}.

\bibitem{SHIFMAN1980493}
\bibinfo{author}{Shifman, M.}, \bibinfo{author}{Vainshtein, A.} \& \bibinfo{author}{Zakharov, V.}
\newblock \bibinfo{title}{Can confinement ensure natural cp invariance of strong interactions?}
\newblock \emph{\bibinfo{journal}{Nuclear Physics B}} \textbf{\bibinfo{volume}{166}}, \bibinfo{pages}{493--506} (\bibinfo{year}{1980}).
\newblock \urlprefix\url{https://www.sciencedirect.com/science/article/pii/0550321380902096}.

\bibitem{DINE1981199}
\bibinfo{author}{Dine, M.}, \bibinfo{author}{Fischler, W.} \& \bibinfo{author}{Srednicki, M.}
\newblock \bibinfo{title}{A simple solution to the strong cp problem with a harmless axion}.
\newblock \emph{\bibinfo{journal}{Physics Letters B}} \textbf{\bibinfo{volume}{104}}, \bibinfo{pages}{199--202} (\bibinfo{year}{1981}).
\newblock \urlprefix\url{https://www.sciencedirect.com/science/article/pii/0370269381905906}.

\bibitem{Zhitnitsky:1980tq}
\bibinfo{author}{Zhitnitsky, A.~R.}
\newblock \bibinfo{title}{{On Possible Suppression of the Axion Hadron Interactions. (In Russian)}}.
\newblock \emph{\bibinfo{journal}{Sov. J. Nucl. Phys.}} \textbf{\bibinfo{volume}{31}}, \bibinfo{pages}{260} (\bibinfo{year}{1980}).

\bibitem{WITTEN1984351}
\bibinfo{author}{Witten, E.}
\newblock \bibinfo{title}{Some properties of o(32) superstrings}.
\newblock \emph{\bibinfo{journal}{Physics Letters B}} \textbf{\bibinfo{volume}{149}}, \bibinfo{pages}{351--356} (\bibinfo{year}{1984}).
\newblock \urlprefix\url{https://www.sciencedirect.com/science/article/pii/0370269384904222}.

\bibitem{Svrcek:2006yi}
\bibinfo{author}{Svrcek, P.} \& \bibinfo{author}{Witten, E.}
\newblock \bibinfo{title}{{Axions In String Theory}}.
\newblock \emph{\bibinfo{journal}{JHEP}} \textbf{\bibinfo{volume}{06}}, \bibinfo{pages}{051} (\bibinfo{year}{2006}).

\bibitem{Conlon:2006tq}
\bibinfo{author}{Conlon, J.~P.}
\newblock \bibinfo{title}{{The QCD axion and moduli stabilisation}}.
\newblock \emph{\bibinfo{journal}{JHEP}} \textbf{\bibinfo{volume}{05}}, \bibinfo{pages}{078} (\bibinfo{year}{2006}).

\bibitem{Choi:2009jt}
\bibinfo{author}{Choi, K.-S.}, \bibinfo{author}{Nilles, H.~P.}, \bibinfo{author}{Ramos-Sanchez, S.} \& \bibinfo{author}{Vaudrevange, P. K.~S.}
\newblock \bibinfo{title}{{Accions}}.
\newblock \emph{\bibinfo{journal}{Phys. Lett. B}} \textbf{\bibinfo{volume}{675}}, \bibinfo{pages}{381--386} (\bibinfo{year}{2009}).

\bibitem{Arvanitaki:2009fg}
\bibinfo{author}{Arvanitaki, A.}, \bibinfo{author}{Dimopoulos, S.}, \bibinfo{author}{Dubovsky, S.}, \bibinfo{author}{Kaloper, N.} \& \bibinfo{author}{March-Russell, J.}
\newblock \bibinfo{title}{{String Axiverse}}.
\newblock \emph{\bibinfo{journal}{Phys. Rev. D}} \textbf{\bibinfo{volume}{81}}, \bibinfo{pages}{123530} (\bibinfo{year}{2010}).

\bibitem{Acharya:2010zx}
\bibinfo{author}{Acharya, B.~S.}, \bibinfo{author}{Bobkov, K.} \& \bibinfo{author}{Kumar, P.}
\newblock \bibinfo{title}{{An M Theory Solution to the Strong CP Problem and Constraints on the Axiverse}}.
\newblock \emph{\bibinfo{journal}{JHEP}} \textbf{\bibinfo{volume}{11}}, \bibinfo{pages}{105} (\bibinfo{year}{2010}).

\bibitem{Cicoli:2012sz}
\bibinfo{author}{Cicoli, M.}, \bibinfo{author}{Goodsell, M.} \& \bibinfo{author}{Ringwald, A.}
\newblock \bibinfo{title}{{The type IIB string axiverse and its low-energy phenomenology}}.
\newblock \emph{\bibinfo{journal}{JHEP}} \textbf{\bibinfo{volume}{10}}, \bibinfo{pages}{146} (\bibinfo{year}{2012}).

\bibitem{Halverson:2017deq}
\bibinfo{author}{Halverson, J.}, \bibinfo{author}{Long, C.} \& \bibinfo{author}{Nath, P.}
\newblock \bibinfo{title}{{Ultralight axion in supersymmetry and strings and cosmology at small scales}}.
\newblock \emph{\bibinfo{journal}{Phys. Rev. D}} \textbf{\bibinfo{volume}{96}}, \bibinfo{pages}{056025} (\bibinfo{year}{2017}).

\bibitem{Demirtas:2018akl}
\bibinfo{author}{Demirtas, M.}, \bibinfo{author}{Long, C.}, \bibinfo{author}{McAllister, L.} \& \bibinfo{author}{Stillman, M.}
\newblock \bibinfo{title}{{The Kreuzer-Skarke Axiverse}}.
\newblock \emph{\bibinfo{journal}{JHEP}} \textbf{\bibinfo{volume}{04}}, \bibinfo{pages}{138} (\bibinfo{year}{2020}).

\bibitem{Hui:2021tkt}
\bibinfo{author}{Hui, L.}
\newblock \bibinfo{title}{{Wave Dark Matter}}.
\newblock \emph{\bibinfo{journal}{Ann. Rev. Astron. Astrophys.}} \textbf{\bibinfo{volume}{59}}, \bibinfo{pages}{247--289} (\bibinfo{year}{2021}).

\bibitem{Read:2014qva}
\bibinfo{author}{Read, J.~I.}
\newblock \bibinfo{title}{{The Local Dark Matter Density}}.
\newblock \emph{\bibinfo{journal}{J. Phys. G}} \textbf{\bibinfo{volume}{41}}, \bibinfo{pages}{063101} (\bibinfo{year}{2014}).

\bibitem{ParticleDataGroup:2024cfk}
\bibinfo{author}{Navas, S.} \emph{et~al.}
\newblock \bibinfo{title}{{Review of particle physics}}.
\newblock \emph{\bibinfo{journal}{Phys. Rev. D}} \textbf{\bibinfo{volume}{110}}, \bibinfo{pages}{030001} (\bibinfo{year}{2024}).

\bibitem{Ema:2016ops}
\bibinfo{author}{Ema, Y.}, \bibinfo{author}{Hamaguchi, K.}, \bibinfo{author}{Moroi, T.} \& \bibinfo{author}{Nakayama, K.}
\newblock \bibinfo{title}{{Flaxion: a minimal extension to solve puzzles in the standard model}}.
\newblock \emph{\bibinfo{journal}{JHEP}} \textbf{\bibinfo{volume}{01}}, \bibinfo{pages}{096} (\bibinfo{year}{2017}).

\bibitem{Calibbi:2016hwq}
\bibinfo{author}{Calibbi, L.}, \bibinfo{author}{Goertz, F.}, \bibinfo{author}{Redigolo, D.}, \bibinfo{author}{Ziegler, R.} \& \bibinfo{author}{Zupan, J.}
\newblock \bibinfo{title}{{Minimal axion model from flavor}}.
\newblock \emph{\bibinfo{journal}{Phys. Rev. D}} \textbf{\bibinfo{volume}{95}}, \bibinfo{pages}{095009} (\bibinfo{year}{2017}).

\bibitem{Maier:2025}
\bibinfo{author}{Maier, R.} \emph{et~al.}
\newblock \bibinfo{title}{Readout of a solid state spin ensemble at the projection noise limit} (\bibinfo{year}{2025}).
\newblock \urlprefix\url{https://arxiv.org/abs/2509.11854}.
\newblock \eprint{2509.11854}.

\bibitem{Capozzi:2020cbu}
\bibinfo{author}{Capozzi, F.} \& \bibinfo{author}{Raffelt, G.}
\newblock \bibinfo{title}{{Axion and neutrino bounds improved with new calibrations of the tip of the red-giant branch using geometric distance determinations}}.
\newblock \emph{\bibinfo{journal}{Phys. Rev. D}} \textbf{\bibinfo{volume}{102}}, \bibinfo{pages}{083007} (\bibinfo{year}{2020}).

\bibitem{XENON:2022ltv}
\bibinfo{author}{Aprile, E.} \emph{et~al.}
\newblock \bibinfo{title}{{Search for New Physics in Electronic Recoil Data from XENONnT}}.
\newblock \emph{\bibinfo{journal}{Phys. Rev. Lett.}} \textbf{\bibinfo{volume}{129}}, \bibinfo{pages}{161805} (\bibinfo{year}{2022}).

\end{thebibliography}

\newpage

\begin{appendices}
\crefalias{section}{appendix}
\numberwithin{equation}{section}

\section{Evaluation of relaxation time scales}
\label{sec:relaxation}

In this section, relaxation time scales of spins are evaluated based on the random noise model
\begin{align}
  H_I(t) &= \lambda f(t) g(t) \sigma_z, \tag{\ref{eq:H_I}}\\
  \Braket{f(t)f(0)} &= e^{-|t|/\tau_c}. \tag{\ref{eq:f2}}
\end{align}
The filter function $g(t)$ denotes the measurement protocol, and its concrete form will be defined below.

\subsection{Ramsey, Hahn Echo, Dynamical Decoupling and Carr-Purcell protocols}
\label{sec:Ramsey}

For comparison purposes, we start with the analysis of the Ramsey, Hahn echo (HE), dynamical decoupling (DD, HE with more $\pi$-pulses), and Carr-Purcell (CP, DD with delays between $\pi/2$-pulses and $\pi$-pulses half of the delays between $\pi$-pulses) protocols for ordinary (e.g. magnetic) fields.
For these protocols, the filter function $g(t)$ is defined as
\begin{align}
  g(t) = \begin{dcases}
    \gamma, & (\text{Ramsey}) \\
    \gamma \left\{
      \Theta(\tau/2 - t) - \Theta(t - \tau/2)
    \right\}, & (\text{HE}) \\
    \gamma \sum_{k=0}^{n-1} \Bigg\{
      \Theta\left[ \left(
        t-k\tilde{\tau}
      \right) \left(
        \frac{2k+1}{2}\tilde{\tau}-t
      \right)\right] \\
      \qquad\quad - \Theta\left[\left(
        t-\frac{2k+1}{2}\tilde{\tau}
      \right) \left(
        (k+1)\tilde{\tau}-t
      \right)\right]
    \Bigg\}, & (\text{DD}) \\
    \gamma \sum_{k=0}^{n-1} \Bigg\{
      \Theta\left[ \left(
        t-k\tilde{\tau}
      \right) \left(
        \frac{4k+1}{4}\tilde{\tau}-t
      \right)\right] \\
      \qquad\quad - \Theta\left[\left(
        t-\frac{4k+1}{4}\tilde{\tau}
      \right) \left(
        \frac{4k+3}{4}\tilde{\tau}-t
      \right)\right] \\
      \qquad\quad + \Theta\left[\left(
        t-\frac{4k+3}{4}\tilde{\tau}
      \right) \left(
        (k+1)\tilde{\tau}-t
      \right)\right]
    \Bigg\}, & (\text{CP})
  \end{dcases}
  \label{eq:filter_ordinary}
\end{align}
where $\gamma=\gamma_e,\gamma_N$ depends on the spin species of use and the unit time $\tilde{\tau}$ for decoupling sequences is related to the observation time as $\tau = n\tilde{\tau}$.

In all of these protocols, the quantum state is initially prepared to be $\rho(t=0) = \ket{+}\bra{+}$, where $\ket{+} = (\ket{g} + \ket{e}) / \sqrt{2}$ is a superposition of the ground and excited states of a qubit, and evolves under $H_I(t)$ for a duration $\tau$.
The quantum state after the free precession is given by
\begin{align}
  \rho(\tau) = \rho(0) - i\int_0^{\tau} dt\, [H_I(t), \rho(t)].
\end{align}
By replacing $\rho(t)$ on the right side by the left side recursively, we obtain
\begin{align}
  \rho(\tau) = \rho(0) + \sum_{n=1}^\infty (-i)^n \int_0^{\tau} dt_1\, \int_0^{t_1} dt_2\, \cdots \int_0^{t_{n-1}} dt_n\, [H_I(t), \rho(0)]_n,
\end{align}
with $[\cdot,\cdot]_n$ defined recursively as
\begin{align}
  [A,B]_n &\equiv [A,[A,B]_{n-1}], \\
  [A,B]_1 &\equiv AB-BA.
\end{align}
By substituting \cref{eq:H_I} on the right side and taking the ensemble average, we obtain
\begin{align}
  \Braket{\rho(\tau)} = \rho(0) + \sum_{n=1}^\infty (-i\lambda)^n \frac{1}{n!} \int_0^{\tau} dt_1 \cdots dt_n\, \Braket{f(t_1) \cdots f(t_n)} g(t_1) \cdots g(t_n) [\sigma_z, \rho(0)]_n,
\end{align}
where we extend the integration range with a compensation factor $1/n!$ using the permutation symmetry of the integral variables.
Note the property of the multipoint function of Gaussian noises $\Braket{f(t_1) \cdots f(t_{2m-1})} = 0$ and
\begin{align}
  \Braket{f(t_1) \cdots f(t_{2m})} = \frac{1}{2^m m!} \sum_\sigma \Braket{f(t_{\sigma(1)}) f(t_{\sigma(2)})} \cdots \Braket{f(t_{\sigma(2m-1)}) f(t_{\sigma(2m)})},
\end{align}
with $\sigma$ running over all permutations of $2m$ integers.
Using this property, the above expression can be rewritten as
\begin{align}
  \Braket{\rho(\tau)} = \rho(0) + \sum_{m=1}^\infty (-i\lambda)^{2m}\frac{1}{m!} \left[
    \frac{1}{2} \int_0^{\tau} dt_1 \, \int_0^{\tau} dt_2\, \Braket{f(t_1) f(t_2)} g(t_1) g(t_2)
  \right]^m [\sigma_z, \rho(0)]_{2m}.
  \label{eq:rho_before_exponentiation}
\end{align}
The integral in the square bracket can be evaluated for each detection protocol as
\begin{align}
  \frac{1}{2} \int_0^{\tau} dt_1\, \int_0^{\tau} dt_2\, \Braket{f(t_1) f(t_2)} g(t_1) g(t_2) = \frac{h_n}{4\lambda^2},
\end{align}
with
\begin{align}
  h_n = \begin{dcases}
    4\lambda^2 \gamma^2 \tau_c \left(
      \tau - \tau_c (1 - e^{-\frac{\tau}{\tau_c}})
    \right), & (\text{Ramsey}) \\
    4\lambda^2 \gamma^2 \tau_c \left(
      \tau - \tau_c (3 + e^{-\frac{\tau}{\tau_c}} - 4e^{-\frac{\tau}{2\tau_c}})
    \right), & (\text{HE}) \\
    4\lambda^2 \gamma^2 \tau_c \left\{
      \tau + \tau_c \left(
        \left( 1-e^{-\frac{\tau}{\tau_c}} \right) \tanh^2 \dfrac{\tilde{\tau}}{4\tau_c} - 2n \dfrac{\sinh\frac{\tilde{\tau}}{2\tau_c}}{\cosh^2 \frac{\tilde{\tau}}{4\tau_c}}
      \right)
    \right\}, & (\text{DD}) \\
    4\lambda^2 \gamma^2 \tau_c \left\{
      \tau - 4\tau_c \left(
        \left( 1-e^{-\frac{\tau}{\tau_c}} \right) \dfrac{\sinh^4 \frac{\tilde{\tau}}{8\tau_c}}{\cosh^2 \frac{\tilde{\tau}}{4\tau_c}} + n\tanh \dfrac{\tilde{\tau}}{4\tau_c}
      \right)
    \right\}. & (\text{CP})
  \end{dcases}
\end{align}
Additionally, the commutation relations of the operators under the initial condition $\rho(0) = \ket{+}\bra{+}$ can be calculated recursively as
\begin{align}
  [\sigma_z, \rho(0)]_{2m} = 2^{2m-1} \sigma_x.
\end{align}
Substituting these expressions into \cref{eq:rho_before_exponentiation}, the time evolution is expressed with $h_n$ defined above as
\begin{align}
  \Braket{\rho(\tau)} \simeq \frac{1}{2} \bm{1} + \frac{1}{2} \sigma_x e^{-h_n},
  \label{eq:def_h_n}
\end{align}
where $\bm{1}$ is the identity matrix.

The dominant noise sources are characterised by a different limit of $\tau_c$ for each protocol.
Firstly, for $\tau_c \gg \tau$, we obtain the following spin decoherence:
\begin{align}
  \Braket{\sigma_x}_\tau \to \begin{dcases}
    e^{-2\lambda^2 \gamma^2 \tau^2}, & (\tau_c \gg \tau,\ \text{Ramsey}) \\
    e^{-\lambda^2 \gamma^2 \tau^3/(3\tau_c)}, & (\tau_c \gg \tau,\ \text{HE}) \\
    e^{-\lambda^2 \gamma^2 \tau^3/(3n^2 \tau_c)}, & (\tau_c \gg \tau,\ \text{DD}) \\
    e^{-\lambda^2 \gamma^2 \tau^3/(12n^2 \tau_c)}, & (\tau_c \gg \tau,\ \text{CP})
  \end{dcases}
  \label{eq:tau_c_infty}
\end{align}
where the ensemble and quantum average of an operator are defined as $\Braket{\cdot}_\tau \equiv \mathrm{Tr} \left[
  \Braket{\rho(\tau)} \cdot
\right]$.
For the Ramsey protocol, these noises with a long correlation time often dominate the dephasing dynamics.
In this case, we identify the dephasing time as $T_2^* = 1/(\sqrt{2}\lambda|\gamma|)$, which gives $\Braket{\sigma_x}_\tau \propto e^{-(\tau/T_2^*)^2}$.
Note that the spin decoherence is suppressed by an additional factor of $\tau/\tau_c$ in this limit for the HE protocol, clearly representing its benefit over the Ramsey protocol.
Secondly, in the limit of small $\tau_c$, we obtain
\begin{align}
  \Braket{\sigma_x}_\tau \to e^{-4\lambda^2\gamma^2\tau_c\tau},~~(\tau_c \to 0)
  \label{eq:tau_c_0}
\end{align}
regardless of the protocol.
When the decoherence dynamics of the HE protocol are dominated by these short-correlation noises, the decoherence time is estimated as $T_2 = 1/(4\lambda^2\gamma^2\tau_c)$ with $\Braket{\sigma_x}_\tau = e^{-\tau/T_2}$.
There is another suppression factor of $12n^2$ for the CP protocol (compared with $3n^2$ for the DD protocol) in the long correlation time limit \cref{eq:tau_c_infty}, which leads to a longer $T_2$ for a larger $n$.
In the limit of $n\to\infty$, it is expected that $T_2$ eventually reaches $\mathcal{O}(T_1)$~\cite{BarGill:2013}, when the longitudinal relaxation dominates, which is beyond the scope of the current noise model.

\subsection{Hybrid-spin decoupling protocol, CP style\label{ymadd}}

Next let us move to the hybrid-spin decoupling protocol discussed in the main text.
The filter function for the hybrid-spin decoupling protocol can be defined as
\begin{align}
  \gHD.
  \tag{\ref{eq:g_HD}}
\end{align}
Note that the hybrid-spin decoupling protocol reduces to the CP protocol in the limit of
\begin{align}
  \tilde{\tau}_N,\tilde{\tau}_e\to \frac{\tilde{\tau}}{2} = \frac{\tau}{2n},~~\tov\to 0,~~\gamma_N \to -\gamma,~~ \gamma_e \to \gamma
  \label{eq:CP_limit}
\end{align}
as can be seen from the comparison with \cref{eq:filter_ordinary}.
For simplicity in later discussions, we assume a natural hierarchy $\tilde{\tau}_N \gg \tilde{\tau}_e$.

Repeating the same calculation as in the previous subsection, the averaged quantum state takes the form $\Braket{\rho(\tau)} \simeq \frac{1}{2} \bm{1} + \frac{1}{2} \sigma_x e^{-h_n}$ with $h_n$ given by
\begin{align}
  h_n =& \dfrac{8\lambda^2 \tau_c^2}{\sinh^2 \left[ \frac{\tilde{\tau}}{2\tau_c} \right]} \left[
    \gamma_N \sinh \left(
      \dfrac{\tilde{\tau}_N}{2\tau_c}
    \right) + \gamma_e \left\{ \sinh \left(
        \dfrac{\tilde{\tau}}{2\tau_c}
      \right) - \sinh \left(
        \dfrac{\tilde{\tau}-\tilde{\tau}_e}{2\tau_c}
      \right)
    \right\}
  \right]^2 \notag \\
  &\qquad\qquad \times \left[
    n\sinh \left(
      \dfrac{\tilde{\tau}}{2\tau_c}
    \right) \exp\left(
      -\dfrac{\tilde{\tau}}{2\tau_c}
    \right) - \sinh \left(
      \dfrac{\tau}{2\tau_c}
    \right) \exp\left(
      -\dfrac{\tau}{2\tau_c}
    \right)
  \right] \notag \\
  &-8n\lambda^2\tau_c^2 \Bigg[
    \gamma_N^2 \sinh\left(
      \dfrac{\tilde{\tau}_N}{2\tau_c}
    \right) \exp\left(
      -\dfrac{\tilde{\tau}_N}{2\tau_c}
    \right) \notag \\
  &\qquad\qquad\quad + \gamma_e^2 \sinh\left(
      \dfrac{\tilde{\tau}_e}{4\tau_c}
    \right) \exp\left(
      -\dfrac{\tilde{\tau}_e}{4\tau_c}
    \right) \left\{
      2 + \exp\left(
        -\dfrac{\tilde{\tau}-\tilde{\tau}_e}{2\tau_c}
      \right) + \exp\left(
        -\dfrac{\tilde{\tau}-\frac{\tilde{\tau}_e}{2}}{2\tau_c}
      \right)
    \right\}\notag \\
  &\qquad\qquad\quad -4\gamma_N\gamma_e \sinh\left(
      \dfrac{\tilde{\tau}_N}{2\tau_c}
    \right) \sinh\left(
      -\dfrac{\tilde{\tau}_e}{2\tau_c}
    \right) \exp\left(
      -\dfrac{\tilde{\tau}-\frac{\tilde{\tau}_e}{2}}{2\tau_c}
    \right)
  \Bigg] \notag \\
  &+ 4\lambda^2\tau_c (\gamma_e^2 \tau_e + \gamma_N^2 \tau_N).
  \label{eq:h_n_HSCP}
\end{align}
As before, the relaxation behaviour of the spins can be extracted from this exponent as $\Braket{\sigma_x}_\tau = e^{-h_n}$.
In the long-correlation limit of $\tau_c \gg \tilde{\tau}_N$, we obtain
\begin{align}
  \Braket{\sigma_x}_\tau \to \exp \left[
    -2\lambda^2 \left(
      \gamma_N \tau_N + \gamma_e \tau_e
    \right)^2 + \mathcal{O}(\tau_c^{-1})
  \right].
\end{align}
The dominant contribution from the first term in parentheses vanishes under the condition \cref{eq:fine-tuning}, when the relaxation behaviour is dominated by the second term of $\mathcal{O}(\tau_c^{-1})$ given for an arbitrary $n$ as
\begin{align}
  \Braket{\sigma_x}_\tau \to \exp \left[
    -\frac{2\lambda^2 \gamma_e^2 \tau_e^2 (3\tau - \tau_N - \tau_e)}{3n^2 \tau_c}
  \right],
\end{align}
which should be compared to the same limit for the CP protocol in \cref{eq:tau_c_infty}.
In the short-correlation limit $\tau_c \ll \tilde{\tau}_e$, on the other hand, we obtain
\begin{align}
  \Braket{\sigma_x}_\tau \to \exp \left[
    -4\lambda^2 \tau_c \left(
      \gamma_N^2 \tau_N + \gamma_e^2 \tau_e
    \right)
  \right],
  \label{eq:tauc_short}
\end{align}
for an arbitrary $n$, which is to be compared with \cref{eq:tau_c_0}.
These two limits demonstrate the resemblance of our hybrid-spin decoupling and CP protocols, which are related through \cref{eq:CP_limit}.
However, due to the two timescales $\tilde{\tau}_N \gg \tilde{\tau}_e$ in our protocol, there is yet another regime, i.e., the intermediate regime $\tilde{\tau}_N \gg \tau_c \gg \tilde{\tau}_e$, where we obtain
\begin{align}
  \Braket{\sigma_x}_\tau \to \exp \left[
    -4\lambda^2 \gamma_N \tau_c \left(
      \gamma_N \tau_N + e^{-\frac{\tov}{\tau_c}} \gamma_e \tau_e
    \right) - \frac{(2n-1)\lambda^2 \gamma_e^2 \tau_e^2}{n^2}
  \right].
  \label{eq:tauc_intermediate}
\end{align}
Note that the condition \cref{eq:fine-tuning} erases the first term in the exponential provided that $\tov \ll \tau_c$, leading to an expression similar to the dephasing behaviour of the Ramsey protocol in \cref{eq:tau_c_infty}.
Since the noise sources in the intermediate regime are DC-like for electron spins, they cause dephasing of the spins during the free precession phases of the electron spin, each of which lasts for $\sim \tilde{\tau}_e$.

The benefits of choosing a larger $n$ can be seen by numerical evaluation of the exponent $h_n$.
In \cref{fig:coherence}a, we show a normalised exponent, $h_n/(2\lambda^2 \gamma_e^2 \tau_e^2)$, as a function of the normalised correlation timescale $\tau_c/\tau_e$ under the fine-tuning condition \cref{eq:fine-tuning}.
The orange, teal, olive, and green lines correspond to the setups with $n=1,2,4$, and $8$, respectively.
The dashed lines show the asymptotic behaviour in the three regions discussed above in the protocol specified by the line colour.
Since the asymptotic behaviour for the short correlation time, $\tau_c \ll \tilde{\tau}_e$, is independent of $n$, we show this with a single black dashed line.
It can be seen that the derived asymptotic behaviours correctly capture all the important features of each curve.
Importantly, we can safely claim that the relaxation is less significant for a larger $n$ from the observation that a curve with a choice of $n$ is always strictly below another curve with a smaller $n$.
The situation is the same as the CP protocol discussed in the previous section, and it is expected that the corresponding relaxation time in principle approaches $\mathcal{O}(T_{1e})$ in the limit of $n\to \infty$.

When there are multiple noise sources with different $\tau_c$, the noise amplitude $\lambda^2$ in each expression should be replaced by the sum of $\lambda^2$ for each source in the corresponding regime of $\tau_c$.
Under this noise model, the $\tau_c$ distribution in real experimental environments could be investigated by comparing relaxation timescales for different protocols.
Firstly, the Ramsey and HE protocols for the nitrogen spin show comparable relaxation time $T_{2N}^* \sim T_{2N} \sim \mathcal{O}(10)\,\si{ms}$~\cite{Waldherr2012}.
This implies that only a small portion of the noise sources possesses $\tau_c \gg \SI{10}{ms}$, which corresponds to the long correlation limit $\tau_c \gg \tilde{\tau}_N$ for a moderate $n\sim \mathcal{O}(1)$.
On the other hand, the large hierarchy $T_{2e}^* \sim \SI{1}{\mu s} \ll T_{2e} \sim \SI{100}{\mu s}$~\cite{Barry:2020,Mizuochi:2009,Wolf:2015} for the electron spin implies that a significant portion of noise sources have correlation time scales $\tau_c \gg \SI{100}{\mu s}$, which is within the intermediate regime.
By combining these considerations with \cref{eq:tauc_intermediate,eq:tauc_short}, we conclude that the hybrid-spin decoupling protocol experiences the decoherence behaviour
\begin{align}
  \decoherence
  \tag{\ref{eq:decoherence}}
\end{align}
where $T_2^*$ and $T_2$ are comparable to $T_{2e}^*\sim \SI{1}{\mu s}$ and $T_{2e}\sim \SI{100}{\mu s}$ for the electron spin.
Note that the numerator of the exponent is determined by $\tau_e$, which is much shorter than the total observation time $\tau=\tau_e+\tau_N$, due to partial cancellation of noise.
This also implies that, similar to ordinary dynamic decoupling protocols, the hybrid-spin decoupling protocol provides a longer coherence time for a larger $n$, which asymptotes to $T_{2e}$.
Since $T_{2e}$ can be made longer at lower temperatures, cryogenic environments are highly beneficial for the proposed hybrid-spin decoupling protocol.

\subsection{Hybrid-spin decoupling protocol, DD style}

For completeness, we also look at the DD version of the hybrid-spin decoupling protocol, which is less effective, since the CP version fully cancels both DC and linear noise, while DD only DC noise. The filter function for the DD version can be defined as
\newcommand{\gHDD}{%
  g(t) = \sum_{k=0}^{n-1} \Bigg\{ \,
    & \gamma_e \Theta\left[\left(
        t-k\tilde{\tau}
      \right) \left(
        k\tilde{\tau}+\tilde{\tau}_N-t
      \right)\right] \notag \\
    & + \gamma_N \Theta\left[ \left(
        t-k\tilde{\tau}-\tilde{\tau}_N-\tov
      \right) \left(
        k\tilde{\tau}+\tilde{\tau}_N+\tilde{\tau}_e+\tov-t
      \right)\right]
  \Bigg\}
}
\begin{align}
  \gHDD.
  \label{eq:g_HDD}
\end{align}
Note that this hybrid-spin decoupling protocol reduces to the DD protocol in the limit of \cref{eq:CP_limit}.
The relaxation of the quantum state is given by \cref{eq:def_h_n} with
\begin{align}
  h_n =& \dfrac{8\lambda^2 \tau_c^2}{\sinh^2 \left[ \frac{\tilde{\tau}}{2\tau_c} \right]} \Bigg[
    \gamma_N^2 \sinh^2 \left(
      \dfrac{\tilde{\tau}_N}{2\tau_c}
    \right) + \gamma_e^2 \sinh^2 \left(
      \dfrac{\tilde{\tau}_e}{2\tau_c}
    \right) \notag \\
  &\qquad\qquad\qquad + 2\gamma_N \gamma_e \sinh \left(
      \dfrac{\tilde{\tau}_N}{2\tau_c}
    \right) \sinh \left(
      \dfrac{\tilde{\tau}_e}{2\tau_c}
    \right) \cosh \left(
      \dfrac{\tilde{\tau}}{2\tau_c}
    \right)
  \Bigg] \notag \\
  &\qquad\qquad \times \left[
    n\sinh \left(
      \dfrac{\tilde{\tau}}{2\tau_c}
    \right) \exp\left(
      -\dfrac{\tilde{\tau}}{2\tau_c}
    \right) - \sinh \left(
      \dfrac{\tilde{\tau}}{2\tau_c}
    \right) \exp\left(
      -\dfrac{\tilde{\tau}}{2\tau_c}
    \right)
  \right] \notag \\
  &-8n\lambda^2\tau_c^2 \Bigg[
    \gamma_N^2 \sinh\left(
      \dfrac{\tilde{\tau}_N}{2\tau_c}
    \right) \exp\left(
      -\dfrac{\tilde{\tau}_N}{2\tau_c}
    \right) + \gamma_e^2 \sinh\left(
      \dfrac{\tilde{\tau}_e}{2\tau_c}
    \right) \exp\left(
      -\dfrac{\tilde{\tau}_e}{2\tau_c}
    \right) \notag \\
  &\qquad\qquad\quad -2\gamma_N\gamma_e \sinh\left(
      \dfrac{\tilde{\tau}_N}{2\tau_c}
    \right) \sinh\left(
      -\dfrac{\tilde{\tau}_e}{2\tau_c}
    \right) \exp\left(
      -\dfrac{\tilde{\tau}}{2\tau_c}
    \right)
  \Bigg] \notag \\
  &+ 4\lambda^2\tau_c (\gamma_e^2 \tau_e + \gamma_N^2 \tau_N),
\end{align}
with which the relaxation behaviour is expressed as $\Braket{\sigma_x}_\tau = e^{-h_n}$.
By taking the long-correlation limit of $\tau_c \gg \tilde{\tau}_N$, we obtain
\begin{align}
  \Braket{\sigma_x}_\tau \to \exp \left[
    -2\lambda^2 \left(
      \gamma_N \tau_N + \gamma_e \tau_e
    \right)^2 + O(\tau_c^{-1})
  \right].
\end{align}
Under the condition \cref{eq:fine-tuning}, the second term of $\mathcal{O}(\tau_c^{-1})$ dominates as
\begin{align}
  \Braket{\sigma_x}_\tau \to \exp \left[
    -\frac{2\lambda^2 \gamma_e^2 \tau_e^2 (3\tau - \tau_N - \tau_e)}{3n^2 \tau_c}
  \right].
\end{align}
In the short-correlation limit $\tau_c \ll \tilde{\tau}_e$, on the other hand, we obtain
\begin{align}
  \Braket{\sigma_x}_\tau \to \exp \left[
    -4\lambda^2 \tau_c \left(
      \gamma_N^2 \tau_N + \gamma_e^2 \tau_e
    \right)
  \right],
\end{align}
while in the intermediate regime, $\tilde{\tau}_N \gg \tau_c \gg \tilde{\tau}_e$,
\begin{align}
  \Braket{\sigma_x}_\tau \to \exp \left[
    -4\lambda^2 \gamma_N \tau_c \left(
      \gamma_N \tau_N + e^{-\frac{\tov}{\tau_c}} \gamma_e \tau_e
    \right) - \frac{2\lambda^2 \gamma_e^2 \tau_e^2}{n}
  \right].
\end{align}
By comparing these expressions with those in the previous subsection, we conclude that this hybrid-spin decoupling version results in similar relaxation behaviours, with up to a factor of $2$ larger exponents under the fine-tuning condition and $\tov\to 0$.

\subsection{Comments on the noise model}

Finally, we comment on the noise sources beyond the noise model adopted here.
\cref{eq:H_I} does not describe situations where electron and nitrogen spin feel different fields.
This kind of situations include noise sources like the dipole interaction from a lattice defect quite close to the NV center, the hyperfine interaction between electron and nitrogen spins, and the longitudinal relaxation effect of the electron spin.
The former has a detrimental effect on the coherence time in any case, hence we consider that with proper diamond growth, this should (on average) not be an issue. The hyperfine interaction is of course rather important, and we consider that continuously illuminating the NV centre during $\tau_N$ will minimise any effect, since the electron spin (and after swapping the nuclear spin) is only in one state~\cite{Maurer:2012}. Finally, the longitudinal relaxation time is the limiting factor for large $n$, and constant illumination should help here too~\cite{Maurer:2012}. The importance of these noise sources should be experimentally determined for each setup. We leave these studies as future work. Moreover, one could consider that this protocol could specifically probe such noise sources as well.

\section{Simulation of the protocol}
\label{sec:simulation}

For the numerical simulations in \cref{fig:coherence} and \cref{fig:noise}, we focus on the effect of the principle of the hybrid-spin decoupling protocol. As such, the swaps are instantaneous, thus $\tov=0$. Moreover, to clearly show the effect in plots, we choose $\gamma_N=-0.5\gamma_e$. Hence, $\tau_N\gg\tau_e$ is not satisfied, but it still provides some insight into the protocol.

The simulated Hamiltonian for each spin-$1$ system is
\begin{align}
    H=hD\left(S_z^2-\frac{S_x^2+S_y^2+S_z^2}{3}\right)+h\gamma_\chi\bm{B}\cdot\bm{S}
\end{align}
where $h$ is Planck's constant, $D$ the zero-field splitting, $S_i$ $(i=x,y,z)$ the spin operators, $\gamma_\chi$ the gyromagnetic ratio of spin $\chi$, and $\bm{B}$ the magnetic field. The latter includes both an offset field, a possible gradient field, and any source of noise. The evolution of the Hamiltonian is calculated with small time steps during the duration of the hybrid-spin decoupling protocol sequence.

The kind of noise depends on the simulation. For white noise, at each time step, the noise is sampled from a normal distribution. As the time step is not infinitely small, in principle there is an upper limit on the frequencies included in this ``white'' noise. However, since the delays used in the simulations are orders of magnitude longer than the time step (which is about $\SI{10}{ps}$), their effect is negligible compared to the lower frequency noise. For pink noise (frequency-dependent noise), each noise spectrum is created with a random amplitude for each frequency sampled from a normal distribution, while the phase is sampled from a uniform distribution. In addition, the amplitude is inversely scaled by the frequency. The time series follows from an inverse Fourier transform of this spectrum. Finally, for fixed noise frequencies with frequencies significantly below the resolution of the frequency spectrum, a time sequence is created by adding $N$ sinuses with random amplitude (normal) and phase (uniform). For each source of noise, the amplitudes are chosen such that the coherence time is within the maximum time we can simulate reasonably. As such, absolute values do not have meaning; instead, the differences are investigated when changing the number of repetitions $n$.

For the coherence time curves in \cref{fig:coherence}b, for each $n$, for $10$ different total interaction times $\tau=\tau_e+\tau_N$, the sequence is simulated. Then, the sequence is repeated $100$ times, where each time the added noise is randomly chosen. As such, the decay due to noise is captured in numerical calculations. This is repeated for all three forms of noise. Note that this does not include other forms of decay, but since here $T_2\ll T_1$, it gives a decent indication of the noise cancellation.

For \cref{fig:noise}c and d, each line represents one simulation of the sequence, with DC noise in \cref{fig:noise}c and pink noise in \cref{fig:noise}d. Moreover, a magnetic-field gradient is added, thus there is a field difference between the two spins. For \cref{fig:noise}e, each sequence is simulated $500$ times, again with different noise each time, and the final result of each sequence is collected. Then, the standard deviation of these $500$ final results is calculated, which gives the uncertainty of a specific sequence. This is repeated for several $n$, giving the relation between $n$ and the uncertainty for a fixed sequence length.

\section{Evaluation of noise effects}
\label{sec:noise}

In this appendix, we evaluate the stochastic noise effect on the result of various measurement protocols.
The noise effect is modelled by the interaction Hamiltonian \cref{eq:H_I}, and the noise profile is characterised by its power spectral density
\begin{align}
  P(\omega) \equiv \lambda^2 \int dt\, e^{i\omega t} \Braket{f(t) f(0)}.
  \tag{\ref{eq:noise_PSD}}
\end{align}
By repeating the same calculation as \cref{sec:relaxation}, the external noise leads to the relaxation behaviour of the quantum state described as
\begin{align}
  \Braket{\sigma_x}_\tau = \exp\left[
    -4 \int_0^\infty \frac{d\omega}{2\pi}\, G(\omega)
  \right],
  \tag{\ref{eq:relaxation_noise}}
\end{align}
with
\begin{align}
  G(\omega) = P(\omega) \int_0^\tau dt_1\,dt_2\, e^{-i\omega(t_1-t_2)} g(t_1) g(t_2) ,
  \tag{\ref{eq:G}}
\end{align}
where we used $P(-\omega) = P(\omega)$ to restrict the integral range to positive $\omega$.

For comparison purposes, we evaluate the noise effects on the Ramsey protocol using the filter function $g(t)$ given in \cref{eq:filter_ordinary}. This results in
\begin{align}
  G(\omega) = \frac{2P(\omega)}{\omega^2} \gamma^2 (1-\cos(\omega\tau)),
  \label{eq:G_Ramsey}
\end{align}
which has asymptotic behaviours
\begin{align}
  \begin{cases}
    G(\omega) \to \gamma^2 \tau^2 P(\omega), & (\omega \ll 1/\tau) \\
    G(\omega) \propto \omega^{-2}. & (\omega \gg 1/\tau)
  \end{cases}
  \label{eq:G_Ramsey_asympt}
\end{align}

\subsection{Hybrid-spin decoupling protocol, CP style}

For the hybrid-spin decoupling protocol, the filter function \cref{eq:g_HD} leads to the expression
\begin{align}
  G(\omega) =& \dfrac{4 P(\omega)}{\omega^2}
  \frac{\sin^2\left[\frac{1}{2} \omega \tau \right]}{\sin^2\left[\frac{1}{2} \omega \tilde{\tau}\right]} \notag \\
  &\times \left[
    \gamma_N \sin \left(
      \dfrac{\omega \tilde{\tau}_N}{2}
    \right) + \gamma_e \left(
      \sin \left(
        \dfrac{\omega \tilde{\tau}}{2}
      \right) - \sin \left(
        \dfrac{\omega (\tilde{\tau} - \tilde{\tau}_e)}{2}
      \right)
    \right)
  \right]^2,
  \label{eq:G_HD}
\end{align}
where the total measurement time $\tau = n\tilde{\tau}$ with $\tilde{\tau} = \tilde{\tau}_N + \tilde{\tau}_e + 2\tov$ involves the operation time of the swap gate, $\tov$, which could affect the noise cancellation.
When the low-frequency expansion for $\omega \ll 1/\tau_N$ is considered, the leading contribution is
\begin{align}
  \GHDLF,
  \tag{\ref{eq:G_HD_LF}}
\end{align}
which clearly shows the cancellation of DC-like noises under the condition \cref{eq:fine-tuning}.
After the cancellation, we check that the leading term of the low-frequency expansion reads
\begin{align}
  \GHDFT.
  ~~
  (\text{fine-tuned})
  \tag{\ref{eq:G_HD_FT}}
\end{align}
Not only the leading term, but also the next-to-leading term of $\mathcal{O}(\omega^2)$ vanishes, resulting in highly-suppressed contributions $\propto \omega^4$. 
The effects of finite $\tov$ will be verified by numerical calculation below.

In \cref{fig:noise}a, we plot the noise contribution $G(\omega)$ as a function of the angular frequency $\omega$ for various protocols, assuming white noise $P(\omega) \propto \omega^0$.
The $y$-axis is normalised as $G(\omega) / (\gamma^2 \tau^2 P(\omega_0))$ with a choice of $\omega_0 = \SI{1}{Hz}$ so that the contributions for the Ramsey protocols approximately become unity for $\omega = \omega_0 \ll 2\pi/\tau_N$ [see \cref{eq:G_Ramsey_asympt}].
The blue dashed, magenta dotted, and orange solid lines correspond to $e$-Ramsey, $N$-Ramsey and to the hybrid-spin decoupling with $n=1$ and $\tov=0$ protocols, respectively.
Hereafter, we use $\chi$-Ramsey to represent the Ramsey protocol for the spin of $\chi=N,e$, and $N$ is an abbreviation of $\ce{^14N}$.
For this estimation, $\tau_N = \SI{3.6}{ms}$ and $|\gamma_e/\gamma_N| = 9.1\times 10^3$ are assumed, resulting in the fine-tuning requirement of $\tau_e \simeq \SI{0.40}{\mu s}$.
Instead of showing the oscillating behaviour for the high-frequency region of the Ramsey protocol lines, we show an upper bound on the contribution $\propto \omega^{-2}$ to demonstrate the noise shape.

From the observation of the Ramsey protocol lines, it can be seen that the white noise contribution is flat for the low-frequency region with $\omega \tau \lesssim 1$, where $\tau=\tau_N$ or $\tau_e$ depending on the setup.
Compared to them, the noise contribution for the hybrid-spin decoupling protocol is suppressed by orders of magnitude for the low-frequency region with $\omega \tau_N \lesssim 1$ or $\omega \lesssim \SI{1}{kHz}$, which shows its effectiveness.
The low-frequency behaviour of these lines can be understood from the $\omega$ expansion shown in \cref{eq:G_Ramsey_asympt,eq:G_HD_FT}.
As discussed above, effects of $\tov$ are negligible for a realistic setup $\tov \sim \SI{1}{\mu s} \ll \tau_N$, as visible in \cref{fig:noise}b.

We show similar plots with varying $n$ in \cref{fig:noise}a as well.
Those shown by dashed straight lines are the predicted low-frequency behaviours for the corresponding hybrid-spin decoupling protocols.
The more effective noise suppression for a larger $n$, $G(\omega) \propto n^{-4}$ as in \cref{eq:G_HD_FT}, is confirmed.
Moreover, the lowest frequency at which the $G(\omega)$ touches the blue dashed line shifts to the higher frequencies proportional to $n$.
From \cref{fig:noise}b, effects of $\tov$ are confirmed for $n=4$ as an example.
In particular, the setup with $\tov = 0.1 \times \tilde{\tau}_N/2 = \tau_N/80$ (dashed line) gives almost the same noise suppression as $\tov=0$ (dotted line, same as in \cref{fig:noise}a), but the setup with $\tov = \tilde{\tau}_N/2 = \tau_N/8$ (dashed-dotted line) causes the modification of $\mathcal{O}(1)$, nullifying a part of the benefits of a large $n$ as indicated by \cref{eq:G_HD_FT} after substituting $\tilde{\tau} = \tilde{\tau}_N + \tilde{\tau}_e + 2\tov$.
Conversely, assuming $\tov\sim \SI{1}{\mu s}$ and $\tau_N=\SI{3.6}{ms}$, there is an effective upper bound $n\lesssim \mathcal{O}(100)$ beyond which $\tov$ starts to degrade the noise cancellation performance.
However, it is expected, at least theoretically, that the relaxation timescale becomes longer for a large $n$ similar to the original CP protocol~\cite{BarGill:2013}, which could allow us to work with even larger choices of $n$.

\subsection{Hybrid-spin decoupling protocol, DD style}

Again for completeness, we look at the DD version. The filter function \cref{eq:g_HDD} leads to the expression
\begin{align}
  G(\omega) = \dfrac{4 P(\omega)}{\omega^2}
  \frac{\sin^2\left[\frac{1}{2} \omega \tau \right]}{\sin^2\left[\frac{1}{2} \omega \tilde{\tau}\right]} \Bigg[&
    \gamma_N^2 \sin^2 \left(
      \dfrac{\omega \tilde{\tau}_N}{2}
    \right) + \gamma_e^2 \sin^2 \left(
      \dfrac{\omega \tilde{\tau}_e}{2}
    \right) \notag \\
    &+ 2\gamma_N \gamma_e \sin \left(
      \dfrac{\omega \tilde{\tau}_N}{2}
    \right) \sin \left(
      \dfrac{\omega \tilde{\tau}_e}{2}
    \right) \cos \left(
      \dfrac{\omega \tilde{\tau}}{2}
    \right)
  \Bigg].
  \label{eq:G_HDD}
\end{align}
Compared with the similar equation for the CP-style case, \cref{eq:G_HD}, the third term in the square bracket manifests the difference.
This affects the low-frequency expansion for $\omega \ll 1/\tau_N$.
The leading contribution is
\begin{align}
  G(\omega) \simeq P(\omega) \left[
    (\gamma_N \tau_N + \gamma_e \tau_e)^2
    + \mathcal{O}(\omega^2)
  \right],
\end{align}
which contains the next-leading order terms of $\mathcal{O}(\omega^2)$ that provides non-vanishing contributions under the condition \cref{eq:fine-tuning},
\begin{align}
  G(\omega) \to \frac{\omega^2 P(\omega)}{4n^2} \gamma_e^2 \tau_e^2 \tau^2~~(\text{fine-tuned}).
\end{align}
Comparison with \cref{eq:G_HD_FT} shows that the suppression of noises as $\omega\to 0$ in the DD version occurs more slowly than in the CP version.
Thus, we only discuss the latter approach in the next section and in the main text.

\section{Sensitivity estimation on axion dark matter}
\label{sec:app_sensitivity}

As an example of the possible applications of the hybrid-spin decoupling protocol, we consider axion dark matter searches with NV centres, which was originally considered in \cite{Chigusa:2023roq,Chigusa:2024psk}.
An axion is a $CP$-odd scalar particle\footnote{
Here, $C$ and $P$ stand for the charge and parity conjugation, respectively.
They should not be confused with the Carr-Purcell (CP) protocol.
} that is a pseudo Nambu-Goldstone boson associated with breaking of an anomalous symmetry.
This kind of particle is motivated as a possible solution to the strong $CP$ problem~\cite{Peccei:1977hh,Weinberg:1977ma,Wilczek:1977pj}, often known as a QCD axion~\cite{PhysRevLett.43.103,SHIFMAN1980493,DINE1981199,Zhitnitsky:1980tq}, or as a trace of the string theory in its low-energy effective theory~\cite{WITTEN1984351,Svrcek:2006yi,Conlon:2006tq,Choi:2009jt,Arvanitaki:2009fg,Acharya:2010zx,Cicoli:2012sz,Halverson:2017deq,Demirtas:2018akl}.
At the same time, the axion is also a well-motivated candidate of dark matter, which is abundant around us and leaves non-trivial signals on detectors.
Since the axion has fermion-dependent coupling with fermion spins, it can be efficiently explored by our hybrid-spin decoupling protocol.
For the sensitivity estimation on the axion dark matter, we adopt the same statistical treatment of the time-sequence data as proposed in \cite{Chigusa:2023roq,Chigusa:2024psk}.
For the derivations below, we use natural units.

When an ultralight axion with mass $m_a\ll \si{eV}$ explains the observed dark matter abundance, its occupation number highly exceeds unity.
Since any quantum mechanical effects of such fields are negligible, the axion is well described by the classical field
\begin{align}
  \atx,
  \tag{\ref{eq:axion_field}}
\end{align}
where $\bm{v}_a$ and $\phi$ are the axion velocity and the oscillation phase, both of which obey some fixed distribution functions~(see, for example, \cite{Hui:2021tkt} for a review).
The coherence of the axion field oscillation is maintained only for the coherence time scale
\begin{align}
  \tau_a \equiv \frac{2\pi}{m_a v_0^2} \sim \SI{4e6}{s} \left(
    \frac{\SI{e-15}{eV}}{m_a}
  \right),
\end{align}
beyond which we start to observe different values of $\bm{v}_a$ and $\phi$.
In the above expression, $v_0\sim 10^{-3}$ is the typical size of the axion velocity when axions are assumed to be virialised.
Moreover, the oscillation amplitude is related to the dark matter energy density as\footnote{
Various local and global measurements determine the local dark matter density~\cite{Read:2014qva}.
The results are still widely spread as $\rho_{\mathrm{DM}} \in (0.2, 0.6)\,\si{GeV/cm^3}$ for global methods and $\rho_{\mathrm{DM}} \in (0.3, 1.5)\,\si{GeV/cm^3}$ for local methods using the Gaia satellite~\cite{ParticleDataGroup:2024cfk}.
In this paper, we use the value in \cref{eq:rho_DM} as a popular choice of the reference value.
}
\begin{align}
  \rho_{\mathrm{DM}} = \frac{1}{2}m_a^2 a_0^2 \sim \SI{0.45}{GeV}.
  \label{eq:rho_DM}
\end{align}
In the following, we will neglect the spatial dependence of the axion field unless necessary, considering that the typical wavelength
\begin{align}
  \lambda_a \equiv \frac{2\pi}{m_a v_0} \sim \SI{e12}{m} \left(
    \frac{\SI{e-15}{eV}}{m_a}
  \right),
\end{align}
is much longer than the dimensions of the experimental setups for mass ranges of our interest.

One of the features of the axion as a pseudo Nambu-Goldstone boson is that its interactions with fermions are given by
\begin{align}
  \mathcal{L}_{\mathrm{int}} = \sum_\chi \frac{C_\chi}{2f_a} (\partial_\mu a) \bar{\chi} \gamma^\mu \gamma_5 \chi,
\end{align}
where $f_a$ is the axion decay constant, $\chi=e,p,n,\dots$ is the index of fermion species, and $C_\chi$ are the model-dependent coefficients, which are $\mathcal O(1)$ for well-motivated axion models~\cite{DINE1981199,Zhitnitsky:1980tq,Ema:2016ops,Calibbi:2016hwq}.
In the non-relativistic limit, this reduces to an interaction Hamiltonian
\begin{align}
  \Hint,
  \tag{\ref{eq:Hint}}
\end{align}
where $\gamma_\chi$ and $\bm{S}_\chi$ are the gyromagnetic ratio and the spin operator of $\chi$, respectively.
The effective magnetic field is related to other parameters as
\begin{align}
  \Beff,
  \tag{\ref{eq:B_eff_chi}}
\end{align}
where $m_\chi$ is the fermion mass and $g_{a\chi\chi} \equiv C_\chi m_\chi / f_a$.
Note that the amplitude of $|\bm{B}_\chi|$ depends on $\chi$ through the independence of the coupling constants $g_{a\chi\chi}$.
Following the discussion in \cite{Chigusa:2024psk}, a nucleus $\chi=\ce{^14N}$ can also be treated similarly if the corresponding coupling constant $g_{aNN}$ is properly defined using the proton and neutron couplings as
\begin{align}
  \frac{g_{aNN}}{m_N} \equiv -\frac{1}{6} \left(
    \frac{g_{app}}{m_p} + \frac{g_{ann}}{m_n}
  \right),
\end{align}
where $m_N$ is the mass of $\ce{^14N}$.

The finite coherence time $\tau_a$ of the axion oscillation is reflected by the two-point function of the effective field,
\begin{align}
  \Braket{B_\chi^z (t) B_\chi^z (0)} =& 
  \left. \int \frac{d\phi}{2\pi}\, \int \frac{d\hat{v}_a}{4\pi}\, 
  B_\chi^z (t) \right|_{\bm{v}_a\to v_0 \hat{v}_a} 
  \left. \int \frac{d\phi'}{2\pi}\, \int \frac{d\hat{v}_a'}{4\pi}\,
  B_\chi^z (0) \right|_{\bm{v}_a\to v_0 \hat{v}_a',\,\phi\to\phi'} \notag \\
  &\times \left[
    \Theta(|t| - \tau_a) + 8\pi^2 \delta(\phi-\phi') \delta(\hat{v}_a - \hat{v}_a') \Theta(\tau_a - |t|)
  \right],
\end{align}
where $\Braket{\cdot}$ denotes the ensemble average, and the second and fourth integrals represent the directional integral over the axion velocities.
Substituting \cref{eq:B_eff_chi} into this definition, this can be simplified as
\begin{align}
  \gamma_\chi^2 \Braket{B_\chi^z (t) B_\chi^z (0)} = \frac{\rho_a v_0^2 g_{a\chi\chi}^2}{3m_\chi^2} \cos(m_a t) \Theta(\tau_a - |t|),
\end{align}
where a factor of $1/3$ arises from the directional average of $(v_a^z)^2$.
Using these properties, the axion-induced signal can be treated in the same footing as the noise effects, using the signal amplitude $\lambda_a$, a normalised random function $f_a(t)$, and the protocol-dependent filter functions $g_a(t)$, as will be defined soon.

In our experimental setup, we repeat the measurements for the total observation time $\tobs$, which corresponds to the $\Nobs \sim \tobs / \tau$ repetitions considering the single sequence time $\tau$.
We label each measurement by the index $j=0,\dots,\Nobs-1$ and the corresponding time $t_j$, and represent the quantum state before the projection measurement by $\rho_j$.
For simplicity of discussion, we focus on a single NV centre system unless otherwise noted; extension to an ensemble of NV centres is straightforward.
Then, $\rho_j$ can be expressed as
\begin{align}
  \rho_j &= \ket{\psi_j} \bra{\psi_j}, \label{eq:rho_j} \\
  \ket{\psi_j} &\equiv \cos\left(
    \frac{\pi}{4} - \phi_j
  \right) \ket{0} + \sin\left(
    \frac{\pi}{4} - \phi_j
  \right) \ket{1}. \label{eq:psi_j}
\end{align}
We define the signal strength as
\begin{align}
  \MyTr{\rho_j \sigma_j^z} = 2\phi_j,
\end{align}
where $\sigma_j^z$ is the Pauli operator acting on the quantum state at $t_j$.
This quantity corresponds to the information read out by the fluorescence measurement.
The axion signal and magnetic-noise contributions to $\phi_j$ are calculated as
\begin{align}
  \phi_j &= \int_{0}^{\tau} dt\, \left(
    \lambda_a f_a(t_j) g_a(t_j) + \lambda f(t_j + t) g(t)
  \right), \label{eq:phi_j}
\end{align}
where the axion signal is correctly represented with the following definitions:
\begin{align}
  \lambda_a &= \sqrt{\frac{\rho_a}{3}} v_0, \\
  \Braket{f_a(t)f_a(0)} &= \cos(m_a t) \Theta(\tau_a - |t|), \\
  g_a(t) &= g(t) \Big|_{\gamma_\chi \to \frac{g_{a\chi\chi}}{m_\chi}}.
\end{align}
The expression of $g_a(t)$ is valid for any detection protocol if the corresponding filter function $g(t)$ is used on the right side.
For the magnetic noise, we consider the spatially uniform noise considered in the previous section.\footnote{
It is possible to also take into account the relaxation effect on the right side of \cref{eq:phi_j}, but we instead empirically take account of them by rescaling the signal strength with a factor of exponential decay.
}

We characterise the signal using the two-point function
\begin{align}
  C_{jj'} \equiv \Braket{\MyTr{\rho_{jj'} \sigma_j^z \sigma_{j'}^z}},
\end{align}
with
\begin{align}
  \rho_{jj'} \equiv \begin{cases}
    \rho_j \otimes \rho_{j'}, & (j\neq j') \\
    \rho_j. & (j = j')
  \end{cases}
\end{align}
Practically, we more often use the PSD for this sequence of data, defined as
\begin{align}
  \PSD{k} \equiv \frac{\tau^2}{\tobs} \sum_{j,j'} e^{2\pi i k (j-j')/\Nobs} C_{jj'},
  \label{eq:PSD_def}
\end{align}
with $k=0,\dots,\Nobs-1$.
By taking the continuum limit, this expression can be rewritten as
\begin{align}
  \PSD{k} \simeq \frac{1}{\tobs} \int_0^{\tobs} dt\, \int_0^{\tobs} dt'\, e^{i\omega_k (t-t')} C(t,t'),
\end{align}
where $\omega_k \equiv 2\pi k/\tobs$ and $C(t,t')$ is a natural interpolation of $C_{jj'}$.

By substituting \cref{eq:rho_j,eq:psi_j,eq:phi_j} into the above definition, the PSD can be expressed as
\begin{align}
  \PSD{k} &= \tau + \SPSD{k} + \BPSD{k}, \\
  \SPSD{k} &\simeq \frac{\lambda_a^2}{\tobs} \int_0^{\tobs} dt\, dt'\, e^{i\omega_k (t-t')} \int_{0}^{\tau} dt_1\, dt_2\, \Braket{f_a(t + t_1) f_a(t' + t_2)} g_a(t_1) g_a(t_2), \\
  \BPSD{k} &\simeq \frac{\lambda^2}{\tobs} \int_0^{\tobs} dt\, dt'\, e^{i\omega_k (t-t')} \int_{0}^{\tau} dt_1\, dt_2\, \Braket{f(t + t_1) f(t' + t_2)} g(t_1) g(t_2),
  \label{eq:BPSD}
\end{align}
where the constant term arises from the terms with $j=j'$.
Independence of the axion signal and the magnetic noise is used to decompose the remaining part of $\PSD{k}$.
It can be seen that the noise contribution $\BPSD{k}$ is connected to the quantities shown in \cref{fig:noise}a as follows.
Using the time translation symmetry, $\Braket{f(t + t_1) f(t' + t_2)} = \Braket{f(t - t' + t_1 - t_2) f(0)}$ is obtained, which can then be used to rewrite the $t$-integral of the two-point function with the noise PSD \cref{eq:noise_PSD} as
\begin{align}
  \BPSD{k} \simeq P(\omega_k) \int_{0}^{\tau} dt_1\, dt_2\, e^{-i\omega_k (t_1-t_2)} g(t_1) g(t_2),
  \label{eq:BPSD}
\end{align}
where we used an approximation $\tobs \gg \tau$.
Compared with \cref{eq:G}, $\BPSD{k} = G(\omega_k)$ is obtained.

The signal contribution to the PSD, $\SPSD{k}$, can also be further simplified.
By performing the $t_1$- and $t_2$-integrals first,
\begin{align}
  \SPSD{k} &\simeq \frac{4\mathcal{A}}{\tobs} \int_0^{\tobs} dt\, dt'\, e^{i\omega_k (t-t')}
  \cos[m_a(t-t')] \Theta(\tau_a - |t-t'|),
\end{align}
where we used an approximation $\tau \ll \tau_a$ to eliminate the $t_1$- and $t_2$-dependence of the Heaviside function.
The coefficient $\mathcal{A}$ is obtained as
\begin{align}
  \mathcal{A} = \dfrac{\lambda_a^2 g_{a\chi\chi}^2}{m_a^2 m_\chi^2} \sin^2 \left(
    \dfrac{m_a \tau_\chi}{2}
  \right),
\end{align}
for the $\chi$-Ramsey with $\chi=N,e$.
By substituting the definition of $\lambda_a$ and using $1/(3\tilde{f}_a) \equiv |g_{aNN}/m_N|$, it can be verified that $\mathcal{A}$ for the Ramsey protocol is the same as the one presented in \cite{Chigusa:2024psk}.
For the hybrid-spin decoupling protocol, on the other hand, we obtain
\begin{align}
  \mathcal{A} =& \dfrac{\lambda_a^2}{m_a^2}
  \frac{\sin^2\left[\frac{1}{2}m_a\tau \right]}{\sin^2\left[\frac{1}{2}m_a\tilde{\tau}\right]} \notag \\
  &\times \left[
    \dfrac{g_{aNN}}{m_N} \sin \left(
      \dfrac{m_a \tilde{\tau}_N}{2}
    \right) + \dfrac{g_{aee}}{m_e} \left(
      \sin \left(
        \dfrac{m_a \tilde{\tau}}{2}
      \right) - \sin \left(
        \dfrac{m_a (\tilde{\tau} - \tilde{\tau}_e)}{2}
      \right)
    \right)
  \right]^2.
  \label{eq:A_HD}
\end{align}
Finally, the $t$- and $t'$-integrals provide us the final expression, which is expressed using $\mathcal{A}$ as
\begin{align}
  \SPSD{k} \simeq \begin{cases}
    \dfrac{2\mathcal{A}}{\tobs \Delta\omega_k^2} \sin^2 \dfrac{\tobs \Delta\omega_k}{2}, & (\tobs < \tau_a) \\[10pt]
    \dfrac{2\mathcal{A}}{\tobs \Delta\omega_k^2} \sin^2 \dfrac{\tau_a \Delta\omega_k}{2} + \dfrac{\tobs-\tau_a}{\tobs\Delta\omega_k} \mathcal{A} \sin\left[
      \tau_a \Delta\omega_k
    \right]. & (\tobs > \tau_a)
  \end{cases}
\end{align}

A few comments on the signal expression $\mathcal{A}$ for the hybrid-spin decoupling protocol are in order.
Firstly, as expected, the expression \cref{eq:A_HD} can be obtained from \cref{eq:G_HD} by replacing $4P(\omega)\to \lambda_a^2$, $\omega\to m_a$, and $\gamma_\chi\to g_{a\chi\chi}/m_\chi$ ($\chi=N,e$).
Secondly, even in the low-frequency regime, it has a leading order contribution of $\mathcal{O}(\omega^0)$,
\begin{align}
  \mathcal{A} \simeq \frac{\lambda^2}{4} \left(
    \frac{g_{aNN}}{m_N} \tau_N + \frac{g_{aee}}{m_e} \tau_e
  \right)^2 + \mathcal{O}(\omega^2),
\end{align}
which does not vanish in general contrary to the ordinary magnetic field effects.
Furthermore, the signal contributions for $n=1$ approach to those in the Ramsey protocol in a limit that one of the axion couplings is turned off; we obtain the $N$-Ramsey for $g_{aee}\to0$ and $e$-Ramsey for $g_{aNN}\to0$ as expected.
For simplicity of discussion, we will focus below on one of these two situations and compare the sensitivities of the corresponding Ramsey protocol and the hybrid-spin decoupling protocol.

What limits the sensitivity to the axion signal is the noise-induced fluctuation of the PSD instead of the PSD.
However, it is not necessary to separately calculate the fluctuation because, as we will briefly show below, any Gaussian noise results in a statistical fluctuation of the same size as its expectation value.
Under the noise model with a Gaussian random function $f(t)$, the statistical fluctuation can be calculated as
\begin{align}
  \Delta \BPSD{k} \equiv \sqrt{
    \Braket{B_k^2} - \Braket{B_k}^2
  },
\end{align}
where $\Braket{B_k} = \BPSD{k}$ and
\begin{align}
  \Braket{B_k^2} \simeq& \frac{\lambda^4}{\tobs[2]} \int_0^{\tobs} dt_1\,dt_2\,dt_3\,dt_4\, e^{i \omega_k (t_{j_1}-t_{j_2}+t_{j_3}-t_{j_4}) / \Nobs} \int_0^\tau dt_1\, dt_2\, dt_3\, dt_4\, \notag \\
  &\times \Braket{f(t_{j_1} + t_1)f(t_{j_2} + t_2)f(t_{j_3} + t_3)f(t_{j_4} + t_4)} g(t_1) g(t_2) g(t_3) g(t_4).
  \label{eq:O_k_sq}
\end{align}
Due to the assumed Gaussianity of the noise, the four-point function of the random function can be decomposed as
\begin{align}
  \Braket{f_1 f_2 f_3 f_4} = \Braket{f_1 f_2} \Braket{f_3 f_4} + \Braket{f_1 f_3} \Braket{f_2 f_4} + \Braket{f_1 f_4} \Braket{f_2 f_3},
  \label{eq:four_point}
\end{align}
with $f_\ell \equiv f(t_{j_\ell} + t_\ell)$ ($\ell=1,2,3,4$).
Noting that the exponential factor in \cref{eq:O_k_sq} is symmetric under exchange of $j_2$ and $j_4$, each of the first and third terms of \cref{eq:four_point} contributes $\BPSD[2]{k}$ to the right-hand side of \cref{eq:O_k_sq}.
On the other hand, the second term results in the contribution
\begin{align}
  \sim \left|
    \frac{\lambda^2}{\tobs} \int_0^{\tobs} dt\, dt'\, e^{i\omega_k(t+t')} \int_0^\tau dt_1\, dt_3\, \Braket{f(t + t_1)f(t' + t_3)} g(t_1) g(t_3)
  \right|^2.
\end{align}
Using the time translation symmetry and Fourier transformation of the two-point function, the $t$- and $t'$-dependence of the expression can be extracted as
\begin{align}
  \int_0^{\tobs} dt\, dt'\, e^{i(\omega_k - \omega)t} e^{i(\omega_k + \omega)t'},
\end{align}
which, in the limit $\tobs \gg \tau$, is proportional to $\delta(\omega_k - \omega) \delta(\omega_k + \omega)$ and thus vanishes unless $\omega = 0$.
Collecting all contributions, we conclude that\footnote{
There are subtleties for the background estimation for the $k=0$ mode.
However, assuming a long enough observation time of $\tobs \sim \mathcal{O}(1) \si{yr}$, no signal is expected in the $k=0$ bin because of the daily and annual modulation of the dark matter wind, which leads to the frequency shift by at least $\sim (\si{day})^{-1}$.
A detailed study of these modulation effects will be an interesting future direction.
}
\begin{align}
  \Delta \BPSD{k}= \BPSD{k}~~(k \geq 1).
\end{align}

Therefore, the total fluctuation of the PSD can be expressed as
\begin{align}
  \Delta \BPSD[\rm total]{k} = \BPSD{k} + \BPSD[\rm proj]{k} + \BPSD[\rm shot]{k},
\end{align}
where the second and third terms on the right side are the contributions from the projection and shot noises, respectively, which have been evaluated in \cite{Chigusa:2024psk}.
For simplicity of discussion, we assume $\BPSD[\rm shot]{k} \ll \BPSD[\rm proj]{k}$, which might be possible with future single-shot readout techniques for ensembles~\cite{Maier:2025}, and focus on the projection noise
\begin{align}
  \BPSD[\rm proj]{0} &\simeq \frac{\sqrt{2}\tau}{M}, \\
  \BPSD[\rm proj]{k\geq 1} &\simeq \frac{\tau}{M},
  \label{eq:BPSD_proj}
\end{align}
where the number of NV centres $M$ is inserted to recover the correct scaling of the sensitivity.\footnote{
A factor of $4$ difference from \cite{Chigusa:2024psk} is due to a different convention for the PSDs $\PSD{k}$.
}
It is interesting to observe the equivalent white noise of the projection noise.
Assuming the white noise profile $P(\omega) = \eta_n^2$ and considering the low frequency limit $\omega\tau\ll 1$, we obtain a flat noise contribution $\BPSD{k} = G(\omega_k) \to \gamma^2\eta_n^2\tau^2$.
By equating this with $\BPSD[\rm proj]{k\geq 1}$, we obtain the magnetic noise amplitude equivalent to the projection noise
\begin{align}
  \eta_n \simeq \SI{9e2}{fT\, Hz^{-1/2}} \left(
    \frac{\SI{3.6}{ms}}{\tau_N}
  \right)^{1/2} \left(
    \frac{10^{12}}{M}
  \right)^{1/2},
  \label{eq:eta_n_proj_N}
\end{align}
for the $N$-Ramsey protocol, and
\begin{align}
  \eta_n \simeq \SI{8}{fT\, Hz^{-1/2}} \left(
    \frac{\SI{0.5}{\mu s}}{\tau_e}
  \right)^{1/2} \left(
    \frac{10^{12}}{M}
  \right)^{1/2},
  \label{eq:eta_n_proj_e}
\end{align}
for the $e$-Ramsey protocol.
The equivalent magnetic noise at its face value is larger for the $N$-Ramsey due to the weaker interaction of the nuclear spins to the magnetic field.
Using the above expressions, the sensitivity to the signal is determined by comparing $\SPSD{k}$ and $\Delta\BPSD[{\mathrm{total}}]{k}$ for a fixed frequency $\omega_k$.

For demonstration of the potential of the hybrid-spin decoupling protocol, we assume a reasonable frequency profile of the external magnetic noise and estimate the sensitivity under its influence.
The noise PSD $P(\omega)$ is assumed to be
\begin{align}
  \sqrt{P(\omega)} = \begin{cases}
    \SI{10}{fT\, Hz^{-1/2}}, & (\omega > \SI{1}{kHz}) \\
    \SI{10}{fT\, Hz^{-1/2}} \left(
      \dfrac{\SI{1}{kHz}}{\omega}
    \right)^{1/2}, & (\omega < \SI{1}{kHz})
  \end{cases}
  \label{eq:P_omega_assumed}
\end{align}
which captures a typical noise behaviour that the white noise background is dominated by $1/f$ noise for the low-frequency region.
Furthermore, comparing the values of $\sqrt{P(\omega)}$ with \cref{eq:eta_n_proj_N,eq:eta_n_proj_e}, it can be seen that the external noise dominates the projection noise for the whole frequency range in setups with $M\gtrsim 10^{16}$ and $M\gtrsim 10^{12}$ for the $N$- and $e$-Ramsey, respectively.
For smaller $M$, the external noise can still become dominant for lower frequencies with larger $1/f$ contributions.

In \cref{fig:darkmatter}a, we compare the sensitivities to the axion-nucleon couplings of the $N$-Ramsey (top black dotted line) and $n=1$ hybrid-spin decoupling (solid) protocols under the assumed external noise \cref{eq:P_omega_assumed}.
The sensitivities are shown in terms of the combined coupling constant $1/\tilde{f}_a \equiv 3|g_{aNN}/m_N|$.
The black lines correspond to $M=10^{12}$ with the total observation time $\tobs=\SI{1}{month}$.
By comparing solid and dotted lines, it can be seen that the hybrid-spin decoupling protocol improves sensitivity whenever the external magnetic noise contribution dominates over the projection noise.
In particular, the enhanced $1/f$ noise effect can be nullified by more precise cancellation at lower frequencies, as seen in \cref{fig:noise}a, recovering the projection noise-limited flat sensitivities obtained in \cite{Chigusa:2024psk}.
Additionally, the prospect sensitivities of the hybrid-spin decoupling protocol with $n=100$ is shown by the black dashed line.
Although a large $n$ also reduces the external noise effect as $\BPSD{k} \propto n^{-4}$ as in \cref{eq:G_HD_FT}, it does not affect the sensitivities since the projection noise $\BPSD[\rm proj]{k}$ always dominates over $\BPSD{k}$ whenever the noise cancellation is effective in the hybrid-spin decoupling protocol.
Instead, the sensitivity improvement from $n=1$ is mainly due to our assumption that a large $n$ makes the coherence time $T_2$ longer by a factor of $100$.
Practically, dependence of $T_2$ on $n$ comes from the detailed noise profile as a function of correlation timescales, so it should be determined experimentally.
In \cref{fig:darkmatter}b, prospects of potential future setups with $M=10^{20}$ and $\tobs=\SI{1}{year}$ are shown to compare with the current best constraints and other theory proporsals.
The conventions for the line color and style are the same as \cref{fig:darkmatter}a.

\begin{figure}[t]
  \centering
  \includegraphics[width=0.6\hsize]{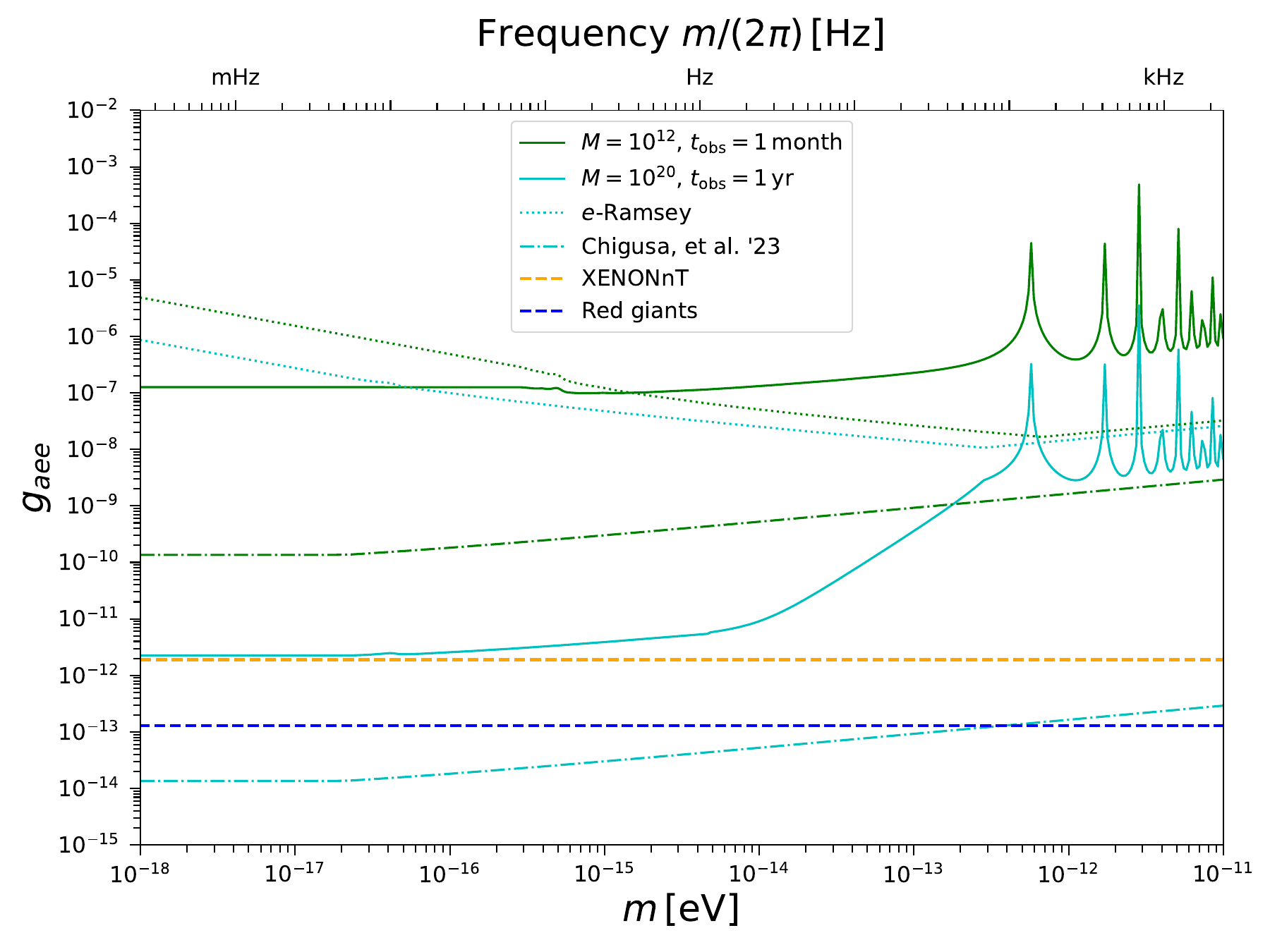}
  \caption{
    Comparison of sensitivities to the axion-electron coupling between the Ramsey (dashed) and hybrid-spin decoupling (solid) protocols under external magnetic noises.
    Also shown by dash-dotted lines are the projection noise-limit sensitivities in \cite{Chigusa:2023roq}.
    The orange and blue dashed lines represent the existing constraints from red giant stars~\cite{Capozzi:2020cbu} and solar axion searches at XENONnT~\cite{XENON:2022ltv}.
  }
  \label{fig:gaee_HD}
\end{figure}

Furthermore, a similar plot for the axion-electron coupling $g_{aee}$ is shown in \cref{fig:gaee_HD}.
Although the hybrid-spin decoupling protocol works fine to cancel the external magnetic noise for the low-frequency region $m \lesssim \SI{1}{kHz}$ as expected, the sensitivity is much lower than the ideal projection noise-limited curve (dotted) obtained in \cite{Chigusa:2023roq}.
This is because, by construction, the hybrid-spin decoupling protocol measures the coupling $g_{aee}$ only for $\tau_e$ out of a single measurement time $\tau \simeq \tau_N + \tau_e \gg \tau_e$.
To achieve a more effective decoupling for the $g_{aee}$ coupling, we need to build a protocol in which a large portion of time is spent for non-trivial dynamics of the electron spin while keeping the overall noise cancellation intact.
This could be an interesting future direction.

\end{appendices}


\end{document}